\documentclass[fleqn,usenatbib]{mnras}

\usepackage[T1]{fontenc}

\DeclareRobustCommand{\VAN}[3]{#2}
\let\VANthebibliography\thebibliography
\def\thebibliography{\DeclareRobustCommand{\VAN}[3]{##3}\VANthebibliography}

\usepackage{graphicx}	
\usepackage{amsmath}	
\usepackage{amssymb}	
\usepackage{newtxtext,newtxmath}
\usepackage{fontawesome}
\usepackage[symbol]{footmisc}

\usepackage{bm}
\usepackage{subcaption}

\newcommand{\lya}{Ly$\alpha$}
\newcommand{\pkt}{$P_{3\mathrm{D},\alpha}$}
\newcommand{\mpcph}{$\rm h^{-1}Mpc$}
\newcommand{\hpmpc}{$\rm h\,Mpc^{-1}$}
\newcommand{\kpcph}{$\rm h^{-1}kpc$}

\newcommand{\alphatitlebold}{\texorpdfstring{$\bm{\alpha}$}{\textbf{α}}}
\newcommand{\git}[2]{\href{https://github.com/#1}{\faGithub}\footnote{\label{#1}#2\url{https://github.com/#1}}}

\usepackage[normalem]{ulem}
\usepackage[dvipsnames]{xcolor}
\def\com#1{{{{#1}}}}

\title[The ACCEL\texorpdfstring{$^2$}{²} project]{The ACCEL\texorpdfstring{$\bm{^2}$}{\textbf{²}} project: simulating Lyman-$\bm{\alpha}$ forest in large-volume hydrodynamical simulations}

\author[S. Chabanier, C. Ravoux et al.]{
Sol{\` e}ne Chabanier$^{1}$\footnotemark[2],
Corentin Ravoux$^{2,3,4}$\footnotemark[2],
Lucas Latrille$^{3}$,
Jean Sexton$^{1}$,
{\' E}ric Armengaud$^{4}$,
\newauthor
\ Julian Bautista$^{3}$,
Tyann Dumerchat$^{3}$,
Zarija Luki{\' c}$^{1}$
\\
$^{1}$Lawrence Berkeley National Laboratory, 1 Cyclotron Road, Berkeley, CA 94720, USA\\
$^{2}$Université Clermont-Auvergne, CNRS, LPCA, 63000 Clermont-Ferrand, France\\
$^{3}$Aix Marseille Universit{\'e}, CNRS/IN2P3, CPPM, Marseille, France\\
$^{4}$Université Paris-Saclay, CEA, IRFU, 91191, Gif-sur-Yvette, France\\
}

\date{Accepted XXX. Received YYY; in original form ZZZ}

\pubyear{2023}

\begin{document}
\label{firstpage}
\pagerange{\pageref{firstpage}--\pageref{lastpage}}
\maketitle

\begin{abstract}
Cosmological information is usually extracted from the Lyman-$\alpha$ (\lya) forest correlations using only either large-scale information interpreted through linear theory or using small-scale information interpreted by means of expensive hydrodynamical simulations. A complete cosmological interpretation of the 3D correlations at all measurable scales is challenged by the need of more realistic models including the complex growth of non-linear small scales that can only be studied within large hydrodynamical simulations. Past work were often limited by the trade off between the simulated cosmological volume and the resolution of the low-density intergalactic medium from which the \lya~signal originates. We conduct a suite of hydrodynamical simulations of the intergalactic medium, including one of the largest \lya~simulations ever performed in terms of volume (640 \mpcph), alongside simulations in smaller volumes with resolutions up to 25 \kpcph\com{, which will be further improved to show resolution convergence in future studies}. We compare the 3D \lya\ power spectra (\pkt) predicted by those simulations to different non-linear models. The inferred \lya~bias and \com{redshift space distortion (RSD)} parameters, $b_\alpha$ and $\beta_\alpha$ are in remarkable agreement with those measured in SDSS and DESI data. We find that, contrary to intuition, the convergence of large-scale modes of the \pkt, which determines $\beta_\alpha$, is primarily influenced by the resolution of the simulation box through mode coupling, rather than the box size itself. Finally, we study the BAO signal encoded in \pkt. For the first time with a hydrodynamical simulation, we clearly detect the BAO signal, however we only marginally detect its damping, associated with the non-linear growth of the structures.
\end{abstract}




\addtocounter{footnote}{-1}
\footnotetext[2]{These two authors equally contributed to this work}

\section{Introduction}
Neutral hydrogen in the intergalactic medium (IGM) scatters light at 1216 \AA, producing characteristic absorption features in the spectra of distant quasars (QSO), dubbed the Lyman-$\alpha$ (\lya) forest. Over the last decade, the Baryon Oscillation Spectroscopic Survey(BOSS)~\citep{Dawson2013} and extended BOSS (eBOSS)~\citep{Dawson2016} through the SDSS (Sloan Digital Sky Survey)~\citep{sdss3,sdss4} have measured the spectra of more than 300,000 high-redshift QSOs allowing for the most precise measurement to date of \lya\ correlations in the IGM.

So far, observational measurements have been conducted on large and small scales independently through the 3D auto-correlation function and the 1D power spectrum, respectively. Large-scale measurements correlate absorption features across different lines of sight and probe matter clustering on scales up to hundreds of \mpcph. They allowed the detection of the Baryon Acoustic Oscillation (BAO) scale at $z=2.3$, providing geometrical constraints on the expansion rate and the angular diameter distance~\citep{Slosar2013,Bautista2017,deSainteAgathe2019,DuMas2020} at a relatively high redshift which has not so far been probed by other observables. More recently, \cite{Cuceu2022} improved these measurements by fitting the "full-shape" auto-correlation function, i.e.~by including smaller scales than BAO analyses, and measured for the first time the Alcock-Paczynski (AP) effect~\citep{Alcock1979}. Small-scale measurements so far focused on correlations within individual quasar lines of sight, through the 1D power spectrum. They probe matter clustering on scales as small as the \mpcph\ scale, unreachable by other tracers~\citep{PalanqueDelabrouille2013,Walther2018,Chabanier2019a,Naim2022,Ravoux2023,Karacayli2023}. The 1D power spectrum is particularly sensitive to the sum of neutrino mass and the nature of dark matter. It has thus been used to put strong constraints on $\sum m_\nu$, the sum of masses of neutrinos~\citep{Seljak2005,PalanqueDelabrouille2015a,PalanqueDelabrouille2015b,PDB2019} as well as several dark matter models, e.g. \com{warm dark matter~\citep{viel_constraining_2005,viel_warm_2013,baur_constraints_2017,Yeche2017,villasenor_new_2023,irsic_unveiling_2024}, fuzzy dark matter~\citep{irsic_first_2017,armengaud_constraining_2017,rogers_strong_2021}} or interacting dark matter~\citep{Dvorkin2014,Xu2018,Garny2018}.

The stage IV dark energy experiment DESI (Dark Energy Spectroscopic Instrument)~\citep{DESI2016,DESI2022}\, as well as other upcoming surveys such as WEAVE~\citep{WEAVE2016}, will considerably improve cosmological measurements from \lya\ observations. DESI, which started its main survey in 2021, is increasing by a factor of 4 the number of high-redshift quasar spectra, going from $\sim$ 300,000 to $\sim$ 1,200,000, and consequently is significantly increasing the number of close quasar pairs, with QSO density increasing from $\sim$ 20 QSO/deg$^2$ in eBOSS~\citep{PalanqueDelabrouille2016} to $\sim$ 60 QSO/deg$^2$ for DESI~\citep{Chaussidon2022}. Therefore, it offers an opportunity to conduct cosmological inference from 3D correlation measurement for the first time through the \lya\ 3D power spectrum (\pkt). Such a measurement would combine cosmological information from both large and small scales, and is expected to significantly improve precision on cosmological parameters and break some parameter degeneracies. For instance, measurement of the sum of neutrino masses would be improved by increasing the statistical power on the Mpc scale (with more pixels separated by a few Mpc than the 1D power spectrum alone) as well as the AP measurement by taking into account even smaller scales than current works. In another example, the 1D power spectrum suffers from major degeneracies between $\sum m_\nu$ and the \com{line broadening due to the IGM thermal state}~\citep{Peeples2010,Kulkarni2015}, which also reduce density fluctuations, but only along the line of sight. 

On the observational side, efforts have already been dedicated to estimating the \pkt~from the data~\citep{FontRibera2018}. First measurements performed on eBOSS data and the forecasted improvement that DESI will provide are given in~\cite{AbdulKarim2023} and~\cite{Belsunce2024}. However, on the theoretical side, we need a robust framework to interpret these complex measurements. On small scales, we have to rely on cosmological hydrodynamical simulations to model the numerous non-linear physical processes that govern the evolution of the baryonic gas in the IGM. A common practice is to run ensembles of simulations to train computationally efficient emulators in order to predict the \lya\ 1D power spectrum in a fast way~\citep{borde_new_2014,Walther2021,Pedersen2021}. 3D correlation modeling is very challenging as we need to simulate large volumes (few hundreds of \mpcph~at least) while resolving the $\sim 100$ kpc~Jeans scale of the low-density IGM. 

The \pkt~computed from such simulations can be fit to analytical models, which add small-scale corrections to the usual redshift-space formula derived from the linear theory of gravitational collapse~\citep{Kaiser1987} for the \lya\ flux field. These small-scale corrections are either generic bias terms defined in a perturbation theory approach~\citep{givans_non-linearities_2022,Ivanov2023}, or empirical functions aimed at  modelling physical phenomena that are at play at those scales~\citep{mcdonald_toward_2003,arinyo-i-prats_non-linear_2015}. However, works so far have been limited by the volume and resolution of such simulations, casting doubts on their interpretation and comparison with data. In particular, $\beta_\alpha$, the \lya~\com{redshift space distortion (RSD)} parameter, was found to be lower in simulations than in observations, e.g. $\beta_\alpha \sim$ 1.4  from 50 \mpcph~\texttt{GADGET-II} simulations~\citep{arinyo-i-prats_non-linear_2015} or $\beta_\alpha \sim$ 1.432  from the 300 \mpcph~\textit{IllustrisTNG} simulations~\citep{Pillepich2019} at $z = 2.3$ while eBOSS DR16 measured $\beta_\alpha = 1.669 \pm 0.071$~\citep{DuMas2020}, and DESI DR1 measured $\beta_\alpha = 1.743_{-0.100}^{+0.074}$~\citep{DESIlya2024}. However, we note that measurements of the $\beta_{\alpha}$ parameter have large uncertainties, as they are degenerate with e.g. the impact of high column density systems which contaminate the \lya~forest.

In order to go beyond past works limited by computational power and set up a robust framework to model 3D correlations of the \lya~forest accurately at the expected level of precision of the data, we constructed the ACCEL$^2$ (ACCELerated expansion of the universe with ACCELerated computing on GPUs) suite of hydrodynamical simulations. It includes one of the largest hydrodynamical simulations of the \lya\ forest to date. In this article, we present the suite of simulations run with the grid-based code \texttt{Nyx} and perform a \pkt~fit based on common analytical models to demonstrate the power of these large-volume simulations and how they improve existing state-of-the-art hydrodynamical simulations. 

The considerable volume reached by our largest simulation makes it possible, for the first time in a hydrodynamical simulation, to characterize the BAO signal encoded in small-wavenumber modes of \pkt ($k < 1$ \hpmpc). Previous studies \com{\citep{Eisenstein2006,Seo2007,kirkby_fitting_2013}} have shown, for both galaxy and \lya~clustering, that the non-linear growth of structures damps the BAO signal. This damping term in the case of \pkt~is actually included routinely in BAO fits (e.g.\com{ in~\cite{DuMas2020}}), but was only measured from an N-body simulation "painted" with the \lya~signal~\citep{Hadzhiyska2023}, and never from hydrodynamical simulations. In this work, we therefore attempt to measure the broadening of the BAO peak from our largest simulation, in order to validate analytical expressions from~\cite{Eisenstein2006}.

The article is structured as followed. Simulations are presented in Sec~\ref{sec:sims}. In particular, we describe the Eulerian code \texttt{Nyx} we use to realize the suite of large-volume hydrodynamical simulations in Sec.~\ref{sec:nyx}, we outline the characteristics of the suite and compare it to state-of-the-art hydrodynamical simulations in Sec.~\ref{sec:suite}, and we present the numerical methods used to extract \pkt~ from the simulations in Sec.~\ref{sec:gimlet}. In Sec.~\ref{sec:models}, we describe the analytical models used in this work. In Sec.~\ref{sec:rsd}, we present \pkt~fit results, and in particular results on the \lya\ bias and RSD parameter. In Sec.~\ref{sec:rsdsplice}, we investigate how these fits are impacted by artificially increasing the physical resolution using the splicing technique. We measure the BAO signal on our largest simulation in Sec.~\ref{sec:bao}. Finally, we conclude and open the road to future works in Sec.~\ref{sec:conclusion}.

\section{The ACCEL\texorpdfstring{$\bm{^2}$}{\textbf{²}} suite of simulations}
\label{sec:sims}
This section presents the suite of \texttt{Nyx} simulations along with the numerical methods to compute \lya\ sightlines and 3D power spectra.
In order to predict the \lya\ forest signal in simulations, the diffuse IGM with typical overdensity $0 \leq \delta \leq 10$ must be modeled.  The evolution of IGM depends on gravity and gas pressure forces which are strongly affected by the reionization model. Following \cite{Chabanier2023} findings, we use the hydrodynamical code \texttt{Nyx}, which we describe in Sec.\ref{sec:nyx}. Then, we present the new suite of simulations and compare it to previous works in Sec.~\ref{sec:suite}.  Finally, the numerical approach to calculating the \pkt~is described in Sec.~\ref{sec:gimlet}.

\subsection{The \texttt{Nyx} code}
\label{sec:nyx}

\texttt{Nyx} is a publicly available\footnote{\url{https://amrex-astro.github.io/Nyx}}, parallel, adaptive mesh, cosmological simulation code that solves the equations of compressible hydrodynamics of baryonic gas coupled with an N-body treatment of the dark matter in an expanding universe \citep{Almgren2013, Sexton2021}. 
\texttt{Nyx}'s hydrodynamics is based on an Eulerian formulation, which is a very efficient approach for the low-density regions of the intergalactic medium. The code uses a second-order (dimensionally-unsplit) piecewise linear (PLM) or piecewise parabolic method \citep[PPM,][]{ppm} to construct the fluxes through the interfaces of each cell. The Poisson equation for self-gravity of the gas and dark matter is solved using a geometric multigrid method.

\texttt{Nyx} is built on the AMReX \citep{AMReX} adaptive mesh refinement (AMR) library and is written in C++. The approach to AMR uses a nested hierarchy of logically-rectangular grids with simultaneous refinement in both space and time. We use MPI to distribute AMR grids across nodes and use logical tiling with OpenMP to divide a grid across threads for multi-core CPU machines (exposing coarse-grained parallelism) and/or CUDA/HIP/DPC++ to spread the work across GPU threads on GPU-based machines (fine-grained parallelism). 

Details of \texttt{Nyx}'s \lya\ forest modeling are given in \citealt{Lukic2015}, but we quickly summarize it here as well. To model the \lya\ forest, \texttt{Nyx} follows the abundance of six species: neutral and ionized hydrogen, neutral, once and twice-ionized helium, and free electrons. For these species, all relevant atomic processes -- ionization, recombination, and free-free transitions are modeled. Heating and cooling source terms are calculated using a sub-cycled approach in order to avoid running the whole code on a short, cooling timescale. It is assumed that all gas elements are optically thin to ionizing photons, such that their ionization state can be fully described by a uniform and isotropic UV+X-ray background radiation field~\citep{Onorbe2017}.

\subsection{The suite of simulations and comparison to existing simulations}
\label{sec:suite}

\begin{table*}
\centering
\begin{tabular}{cccccc}
\hline
Name & Box size  [Mpc/h] & Box size [Mpc] & Number of cells & Physical resolution [kpc/h] \\
\hline
\textit{ACCL2\_L160R100} & 160 & 237 &$1536^3$ & 100  \\
\textit{ACCL2\_L160R50} &  160 & 237 &$3072^3$ & 50  \\
\textit{ACCL2\_L160R25} &  160 &  237 &$6144^3$ & 25  \\
\textit{ACCL2\_L320R100} &  320 & 474 &$3072^3$ & 100  \\
\textit{ACCL2\_L320R50} &  320 &  474 & $6144^3$ & 50  \\
\textit{ACCL2\_L640R100} &  640 & 948 & $6144^3$ & 100  \\
\hline
\end{tabular}
\caption{The set of simulations with the simulation name, the box size in \mpcph~and in Mpc, the number of gas cells, and the equivalent physical resolution in \kpcph. The physical resolutions (precisely 104, 52, and 26 \kpcph) are approximated for clarity and conciseness.}
\label{tab:sims}
\end{table*}

\begin{figure*}
    \centering
    \includegraphics[width = 1.0\textwidth]{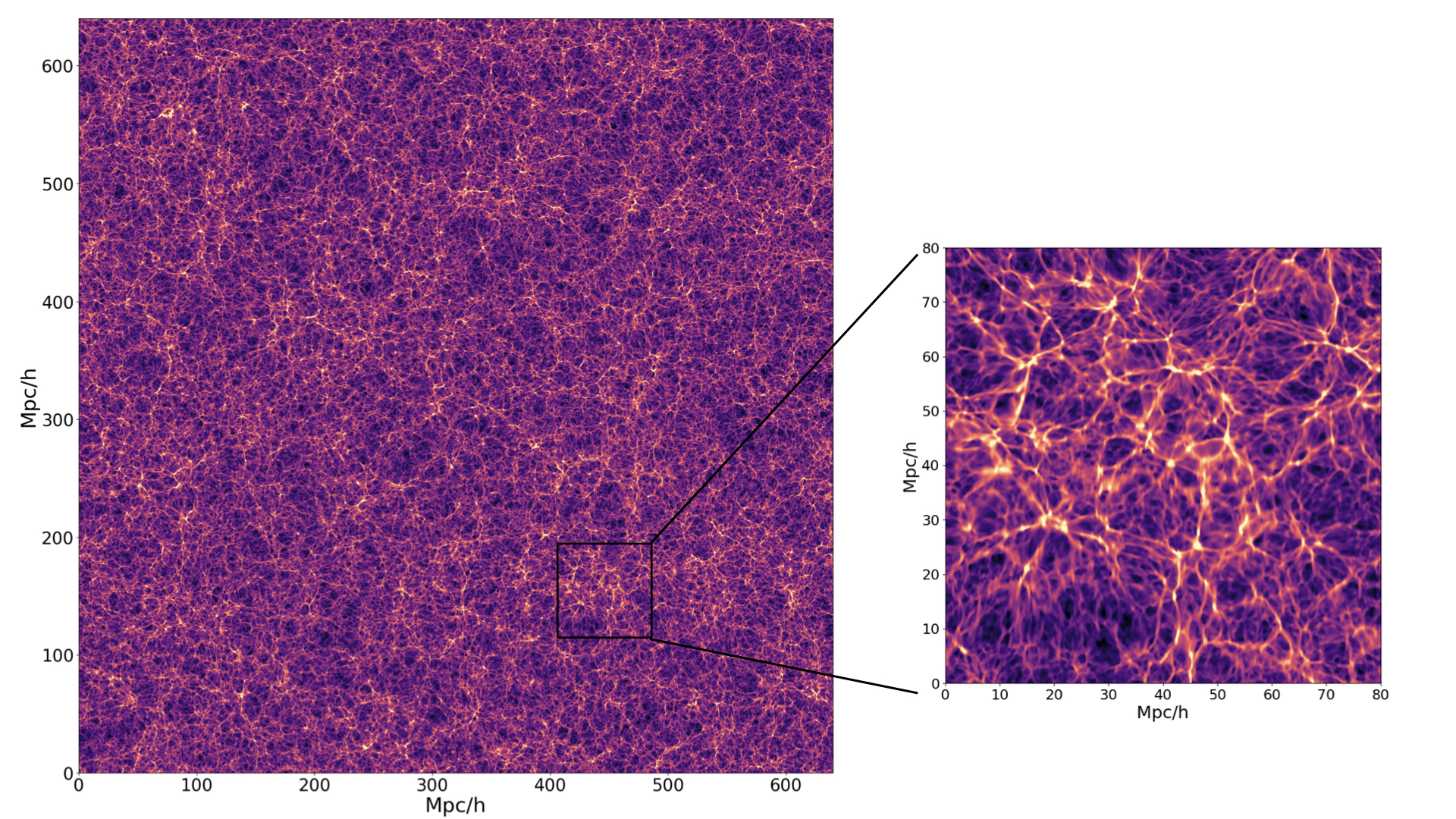}
    \caption{Slice showing the baryon density through the largest simulation, \textit{L640R100}, with 640 \mpcph\ on each side and $6144^3$ cells at $z=2$. The left side shows the whole simulated box, while on the right we show a randomly selected zoom-in region of 80 \mpcph\ width, which is approximately the size probed by current state-of-the-art \lya\ hydrodynamical simulations.}
    \label{fig:sim_box}
\end{figure*}

All simulations are initialized at $z = 200$, using Zel’dovich approximation~\citep{zeldovich1970}. Transfer functions were generated using both analytical approximation~\citep{Einsenstein1999} and the \texttt{CLASS}~\citep{class2011} Boltzmann solver. We use cosmological parameters following Planck 2016~\citep{Planck2015}: $\Omega_b = 0.0487$, $\Omega_m = 0.31$, $H_0 = 67.5$, $n_s = 0.96$ and $\sigma_8 = 0.83$.

We produced a suite of 6 simulations with box size, $L$, ranging from 160 to 640 \mpcph\ , and a number of resolution elements, $N_{c}$, ranging from $1536^3$ to $6144^3$, which translates into resolutions ranging from 25 to 100 \kpcph. We use a uniform grid approach, which works well for low-density regions. Table~\ref{tab:sims} presents a summary of the simulation characteristics. The main simulations are \textit{ACCL2\_L160R25} and \textit{ACCL2\_L640R100}, and we will use \textit{L160R25} and \textit{L640R100} notations in the text for conciseness. The former has the smallest volume of our suite but also has the highest resolution, 25 \kpcph. \com{This has been shown to be an appropriate resolution to study the \lya\ forest down to scales as small as the Mpc \citep{Lukic2015,Walther2021,Chabanier2023}. However, as pointed out in previous studies~\citep{bolton_resolving_2009,Chabanier2023,doughty_convergence_2023}, we note that the required resolution varies as a function of redshift, and that 25 \kpcph~might not necessarily be enough to obtain a fully converged \lya~field for all the redshift considered. The latter simulation has the largest volume, 640 \mpcph, larger than most current state-of-the-art hydrodynamical simulations but has the coarsest resolution.} Intermediate simulations are used to study box size and resolution effects. For each simulation, we have snapshots at z = 2.0, 2.6, 3.0, 3.6, 4.0 and 5.0.

Our suite of simulations considerably improves existing \lya\ simulations not only by exceeding volumes probed but also by increasing the physical resolution in the low-density regime, from which originates the \lya\ forest signal. A visual representation of the baryon density obtained from our largest simulation is given in Fig.~\ref{fig:sim_box}.

For instance, when comparing to the Sherwood suite of simulations (160 \mpcph, 80 \kpcph\ resolution) dedicated to \lya\ studies~\citep{Sherwood}, with \textit{L640R100} we increase the volume of the largest Sherwood simulation by a factor of 64 while keeping an equivalent physical resolution. With \textit{L160R25}, we improve the physical resolution by a factor of 4 while keeping the same cosmological volume. We also probe a larger volume than the MilleniumTNG simulation~\citep{MilleniumTNG} (500 \mpcph, 115 \kpcph\ resolution) by a factor of 2 with a slightly better physical resolution in the intergalactic medium. It is important to keep in mind that the MilleniumTNG simulations model the galaxy formation by including star formation, stellar population evolution, and chemical enrichment following supernovae (SN), supermassive blackhole formation, and galactic feedbacks (SN and blackholes). However, those models have to be included via subgrid models that rely on relatively ad-hoc subgrid-free parameters commonly calibrated on astrophysical observables. The choice of observables on which those are calibrated lead to significant variations on different cosmological observables~\citep{Chisari2019}, in particular couples of percent on the \lya\ P1D~\citep{Chabanier2020b}. During the realization of our study, we noted that the FLAMINGO project~\citep{Schaye2023} realized very large hydrodynamical simulations (700 \mpcph\ with 200 \kpcph\ resolution, and 2000 \mpcph\ with 400 \kpcph\ resolution). Their simulations are probing larger volumes but with a resolution that is significantly lower than our simulations, and are thus not adapted to the study of \lya~forest according to the findings in~\cite{Lukic2015} and~\cite{Chabanier2023}.


\subsection{Extraction of Ly\alphatitlebold\ power spectra}
\label{sec:gimlet}

This section describes the simulation output data and how it is processed to produce synthetic \lya\ forest sightlines and \pkt. All the \lya\ fields and \pkt~ computations are performed with the \texttt{gimlet} post-processing software (see, for instance, \citealt{Briesen2016}). 

For each simulation box, we take skewers along the three simulation axes, keeping periodic boundary conditions with the rays passing through all cell centers. Then, we compute the (normalized) transmitted flux $F$ at every pixel, with $F = e^{-\tau}$ where $\tau$ is the optical depth for \lya\ photon scattering. The latter is defined as
\begin{equation}
\tau_\nu = \frac{\pi e^2}{m_e c}f_{12}\int \frac{n_{\rm HI}}{\Delta_{\nu_{D}}}\phi_\nu dr,
\end{equation}
where $\nu$ is the observed frequency, $e$ is the electron charge,  $c$ is the speed of light, $f_{12}$ is the oscillator strength for the \lya\ resonance transition, $n_{H_I}$ is the neutral hydrogen density, $\Delta \nu_{D} = (b_T/c)\nu_0 = (\sqrt{2 k_B T / m_H} / c)\nu_0 $ is the Doppler width with $b_T$ the Doppler parameter, $m_H$ the mass of hydrogen, $k_B$ the Boltzmann constant and $\phi_\nu$ is the line profile. 

In general, the line profile is a Voigt profile, but we use the Doppler profile instead for several reasons. The Doppler profile is equivalent to the Voigt one for lines whose maximal optical depth is less than 1000. We are only interested in \lya\ forest systems with optical depths at line center of less than 10. For Lyman Limit Systems (LLS) and Damped \lya\ systems (DLAs), our simulations are not designed to compute the correct density and temperature in any case: the HI density in these systems should have self-shielding corrections, which cannot be evaluated properly without coupled radiative transfer-hydrodynamics in the simulations. If we were to use Voigt profiles with these high column density systems, the damping wings would not only be inaccurate, but those errors would then contaminate nearby regions. The Doppler profile is
\begin{equation}
\phi_\nu = \frac{1}{\sqrt{\pi}} \exp \left [ -\left( \frac{\nu - (1-\frac{\nu_{\rm ||}}{c})\nu_0}{\Delta \nu_D} \right)^2 \right].
\end{equation}
In velocity space, peculiar velocities modify the optical depth by shifting the absorption positions and broadening the lines. Thus, in redshift space, we have
\begin{equation}
    \tau_v = \frac{\pi e^2 f_{12}\lambda_0}{m_e c H}\int \frac{n_{\rm HI}}{v_{\rm th}} \exp\left[ - \left( \frac{v-v'-v_{\rm ||}}{v_{\rm th}}, \right)^2\right] dv'
\end{equation}
where $\lambda_0$ is the rest-frame \lya wavelength, $H$ is the Hubble expansion rate at the given redshift\com{, and $v_{\rm th}$ is the Doppler parameter (also named thermal velocity)}. For each output, we use $N_{\rm cells}^2\times 3$ lines of sight, where $N_{\rm cells}$ is the number of resolution elements per dimension.

We define the flux perturbations as
\begin{equation}
    \label{eq:flux_contrast}
    \delta_F = \frac{F}{\left< F\right >}-1\,,
\end{equation}

\noindent where $\left< F\right >$ is the transmitted flux fraction averaged over all skewers. Then, the three-dimensional \lya\ flux power spectrum, $P_{\rm 3D,\alpha} (k,\mu)$ is obtained by taking the average of the squared norm of the Fourier transform of $\delta_F$ and averaging over $k$, the norm of the Fourier mode, and $\mu = k_\parallel/k$, the cosine of the angle between the mode and the line-of-sight considered. 

We average the power spectrum in 4 linear-space bins in $|\mu|$. The wavenumbers $k$ (in \hpmpc) are defined equally for each $\mu$ bin by the \texttt{gimlet} software. It is taken as the linear array between $\pi / L$ and $\pi N / L$, with $N/2$ bins, where $N$ is the number of cells, and $L$ the box size (in \mpcph) for the simulation considered. For each $(k,\mu)$ bin of a power spectrum, we associate a statistical uncertainty $\sigma(k,\mu) = P_{3\mathrm{D},\alpha} / \sqrt{\left(N(k,\mu)\right)}$, where $N(k,\mu)$ is the number of Fourier modes used for the averaging in this mode.

All power spectra are computed as the average of the power spectra computed along each of the three axes of the simulation. For all outputs, we normalize the average transmitted flux fraction such that:

\begin{equation}
    \label{eq:mean_flux}
    \overline{F} = \exp\left(-\tau_{\mathrm{eff}}\right) = \exp\left(- 0.0025 \times (1 + z)^{3.7}\right) \,,
\end{equation} 

\noindent which is in agreement with the redshift evolution found in high-resolution data~\citep{Bolton2014,Walther2019}.

\subsection{Power spectrum on the ACCEL\texorpdfstring{$\bm{^2}$}{\textbf{²}} simulation grid}
\label{sec:p3d}

\begin{figure*}
    \centering
    \includegraphics[trim=0cm 0cm 0cm 0cm, clip=true,width = 1.0\textwidth]{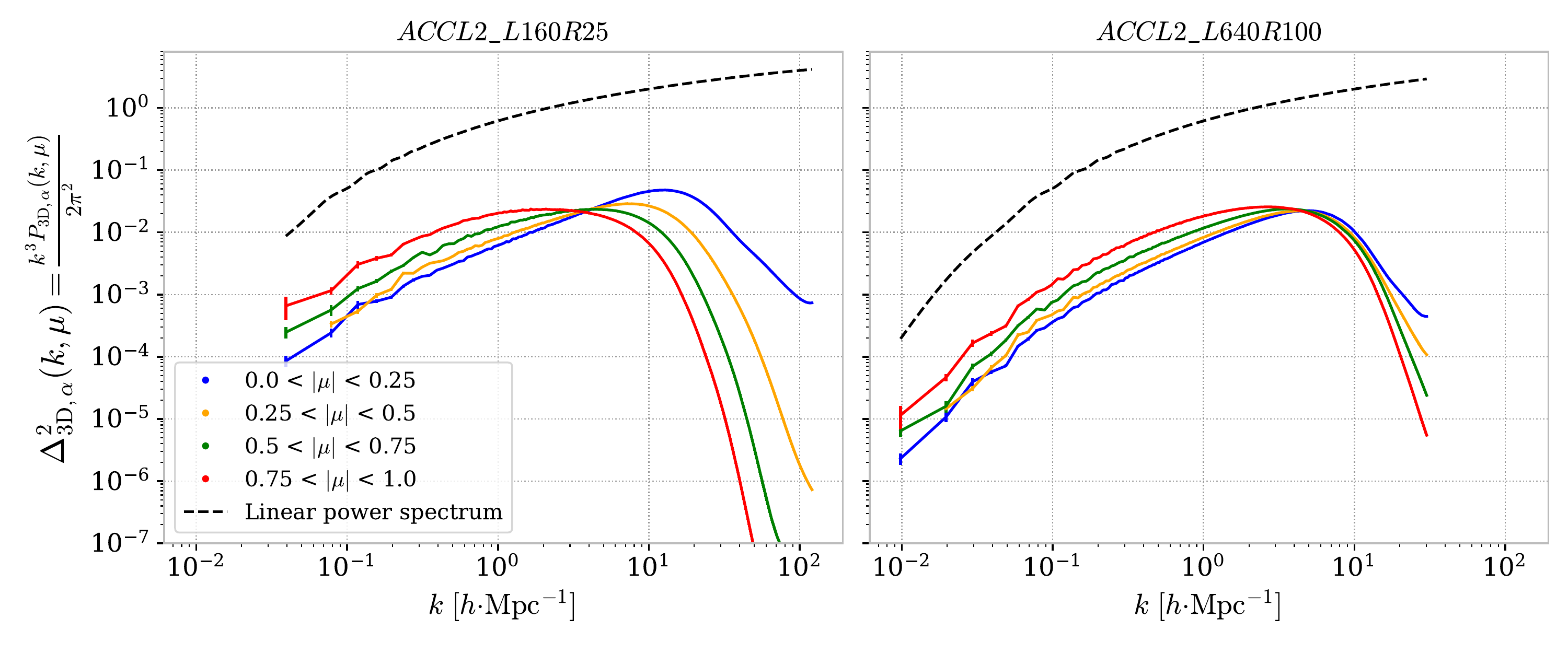}
    \caption{Dimensionless three-dimensional power spectrum ($\Delta_{3\mathrm{D}, \alpha}^2 = k^{3} P_{3\mathrm{D}, \alpha} / 2 \pi^2$) for our most resolved simulation \textit{L160R25} (left) and for our largest simulation \textit{L640R100} (right). Both \pkt~are represented as a function of their wavenumber range (in \hpmpc) for different $\mu$ bins, at redshift $z=2.0$.\com{ As an illustration, the linear matter power spectrum obtained with the \texttt{CLASS}~\citep{class2011} Boltzmann solver and used for the fit is represented in dashed black line in the same wavenumber range.}}
    \label{fig:p3d_plot}
\end{figure*}

The three-dimensional power spectrum of our most resolved simulation \textit{L160R25} and our largest one \textit{L640R100} are represented at redshift $z=2.0$ in Fig.~\ref{fig:p3d_plot}.

For \textit{L160R25}, the resolution is four times smaller than the IGM Jeans length, the scale where gas starts to be pressure supported against gravitational collapse by its temperature, which is sufficient to resolve the small-scale structure of the gas. We measure a \pkt~profile similar to the one found in~\cite{arinyo-i-prats_non-linear_2015}, but with a more converged measurement at small scales, especially for the transverse modes. The \textit{L640R100} simulation measures \pkt~at larger scales ($k \lesssim 0.02$~\hpmpc). However, the resolution of the simulation is insufficient to correctly model the Jeans smoothing and thermal broadening. Indeed, as shown in \com{previous studies~\citep{Lukic2015,Chabanier2023}}, we need a 20 \kpcph~resolution to have \pkt~converged at the level of 2-4\% for wavenumbers up to tens of \hpmpc.

We remind here the physical interpretation for the shape of \pkt~for different $\mu$ bins in \textit{L160R25}. Jeans smoothing for baryons is directly taken into account in our hydrodynamical simulations: the baryonic matter density is smoothed in all directions, imposing an isotropic cut-off on \pkt. The thermal broadening is specific to the \lya~forest, and by definition, it is visible only along the lines-of-sight. Consequently, this broadening impacts the \pkt~in an anisotropic way: the effect is maximum for $|\mu| = 1$ and vanishes transversely ($|\mu| \sim 0$). 

Both effects impact \pkt~at small scales ($k\gtrsim 2$~\hpmpc), but the cut-offs related to thermal broadening and Jeans smoothing start for different $k$. Indeed, the thermal broadening suppression starts at scales $k \sim 2$~\hpmpc. Conversely, Jeans smoothing imposes a power suppression that starts at smaller scales ($k \sim 10$~\hpmpc). The near-transverse $\mu$ bin ($|\mu| < 0.25$) yields a measurement of Jeans smoothing while the near-radial $\mu$ bin ($0.75 < |\mu| < 1$) witnesses the combined effect of Jeans and thermal smoothing, and is dominated by thermal broadening. Between both $\mu$ bins, the transition is progressive.

At larger scales ($k \sim 0.1$~\hpmpc), the difference between $\mu$ bins is mainly driven by the linear RSD Kaiser equation\com{, as it is illustrated in Fig.~\ref{fig:p3d_plot} comparing the linear matter power spectrum and \pkt.}

\section{Analytical Modeling of the 3D power spectrum}
\label{sec:models}

We parameterize \pkt~along the work of~\citet{mcdonald_toward_2003,arinyo-i-prats_non-linear_2015}. Those models were used in more recent \pkt~estimation from simulation~\citep{givans_non-linearities_2022}. The principle is to correct the Kaiser formula developed in the linear framework by a non-linear term $D(k, \mu)$ such that:

\begin{equation}
    \label{eq:model_p3d}
    P_{\mathrm{model},\alpha}(k, \mu)=b_{\alpha}^2\left(1+\beta_{\alpha} \mu^{2}\right)^{2} P_{\mathrm{m}}(k) D(k, \mu)\,.
\end{equation}

The linear terms are $P_{\mathrm{m}}$, the linear matter power spectrum at the considered redshift (computed with the \texttt{CLASS}~\citep{class2011} Boltzmann solver), $b_{\alpha}$ is the \lya~bias, and $\beta_{\alpha}$ is the \lya~RSD parameter. They are theoretically following~\cite{Seljak2005}:

\begin{equation}
    \label{eq:biases}
    \begin{aligned}
    b_{\alpha} &=  \left.\frac{\partial\delta_{F}}{\partial\delta_{\mathrm{m}}}\right\vert_{\delta_{\mathrm{m}} = 0}\,, \\
    \beta_{\alpha} &= \frac{b_{\eta}f}{b_{\alpha}}\,, \\
    b_{\eta} & =  \tau_{\alpha} \left.\frac{\partial\delta_{F}}{\partial\tau_{\alpha}}\right|_{\tau=0}\,,
    \end{aligned}
\end{equation}

\noindent where $\delta_{F}$ is the \lya~contrast, $\delta_{\mathrm{m}}$ is the matter density contrast, $f$ is the logarithmic growth rate of linear perturbations, $b_{\eta}$ is the \lya~velocity bias, and $\tau_{\alpha}$ the \lya~optical depth. Here, all expressed quantities $\left(P_{\mathrm{m}}, b_{\alpha}, \beta_{\alpha}, D\right)$ also depend on redshift. 

The $D(k, \mu)$ term is used to parameterize non-linearities of \pkt~at small scales and must be equal to unity at large scales. At large scales ($k\lesssim 1$~\hpmpc) the \lya~physics is then entirely driven by the $b_{\alpha}$ and $\beta_{\alpha}$ terms, as shown on Fig.~\ref{fig:p3d_plot}. 

The bias $b_{\alpha}$ gives the amplitude of \pkt~with respect to the matter power spectrum, as shown in Fig.~\ref{fig:p3d_plot}. Considering the definition of the \lya~contrast $\delta_F$ in Eq.~\ref{eq:flux_contrast}, the bias $b_{\alpha}$ is negative. Indeed, a \lya~over-absorption ($F < \langle F \rangle$) yields a negative value of $\delta_F$ but corresponds to a matter over-density. The RSD term $\beta_{\alpha}$ is directly proportional to the logarithmic growth rate of structures $f$ and is the source of the dependence of \pkt~as a function of $\mu$ on large scale, as seen in Fig.~\ref{fig:p3d_plot}. 

To model the non-linear contribution $D(k, \mu)$, a first model was developed in~\citet{mcdonald_toward_2003} and rewritten in~\citet{arinyo-i-prats_non-linear_2015} as

\begin{equation}
    \label{eq:model_p3d_nl_0}
    D_{0}(k, \mu)=\exp \left\{\left(\frac{k}{k_{\mathrm{nl}}}\right)^{a_{\mathrm{nl}}}-\left(\frac{k}{k_{\mathrm{p}}}\right)^{a_{\mathrm{p}}}-\left(\frac{k \mu}{k_{\mathrm{v0}}\left(1+k / k_{\mathrm{v1}}\right)^{a_{\mathrm{v1}}}}\right)^{a_{\mathrm{v0}}}\right\}\,,
\end{equation}

\noindent where $k_{\mathrm{nl}}$, $a_{\mathrm{nl}}$, $k_{\mathrm{p}}$, $a_{\mathrm{p}}$, $k_{\mathrm{v0}}$, $k_{\mathrm{v1}}$, $a_{\mathrm{v0}}$, and $a_{\mathrm{v1}}$ are free parameters. The three terms in the exponential correspond respectively to the effects of non-linear growth, Jeans smoothing suppression, and the associated effect of thermal broadening and non-linear peculiar velocities along the line-of-sight. This model is not optimal to describe the largest scales, as it gives a $D(k, \mu)$, which does not quickly converge to unity for large scales ($k\lesssim 1$~\hpmpc).

A second model developed in~\citet{arinyo-i-prats_non-linear_2015} uses perturbation theory predictions to reduce the number of free parameters and includes the matter power spectrum $P_{\mathrm{m}}$:

\begin{equation}
    \label{eq:model_p3d_nl_1}
    \begin{split}
    D_{1}(k, \mu)= \exp \Bigg\{  &\left[q_{1} \frac{k^{3} P_{\mathrm{m}}(k)}{2 \pi^{2}} +q_{2} \left(\frac{k^{3} P_{\mathrm{m}}(k)}{2 \pi^{2}} \right)^{2}\right] \times \\
     & \left. \left[1-\left(\frac{k}{k_{\mathrm{v}}}\right)^{a_{\mathrm{v}}} \mu^{b_{\mathrm{v}}}\right] - \left(\frac{k}{k_{\mathrm{p}}}\right)^{2}\right\}\,,
    \end{split}
\end{equation}

\noindent where $q_{1}$, $q_{2}$, $k_{\mathrm{v}}$, $a_{\mathrm{v}}$, $b_{\mathrm{v}}$, and $k_{\mathrm{p}}$ are free parameters. Here, the non-linear growth corrections at different orders are controlled by the parameters $q_{1}$ and $q_{2}$. In \cite{arinyo-i-prats_non-linear_2015}, the authors also consider this model with a fixed value $q_{2} = 0$. The Jeans smoothing is ruled by $k_{\mathrm{p}}$. Finally, the thermal broadening is handled by $(k_{\mathrm{v}},a_{\mathrm{v}},b_{\mathrm{v}})$.

Both models describe the physics of Jeans smoothing, non-linear structure growth and thermal broadening responsible for the crossing of curves for different $\mu$ bins in Fig.~\ref{fig:p3d_plot}.

\section{Fits results on simulations}
\label{sec:rsd}
\begin{figure*}
    \centering
    \includegraphics[trim=0cm 0cm 0cm 0cm, clip=true,width = 1.0\textwidth]{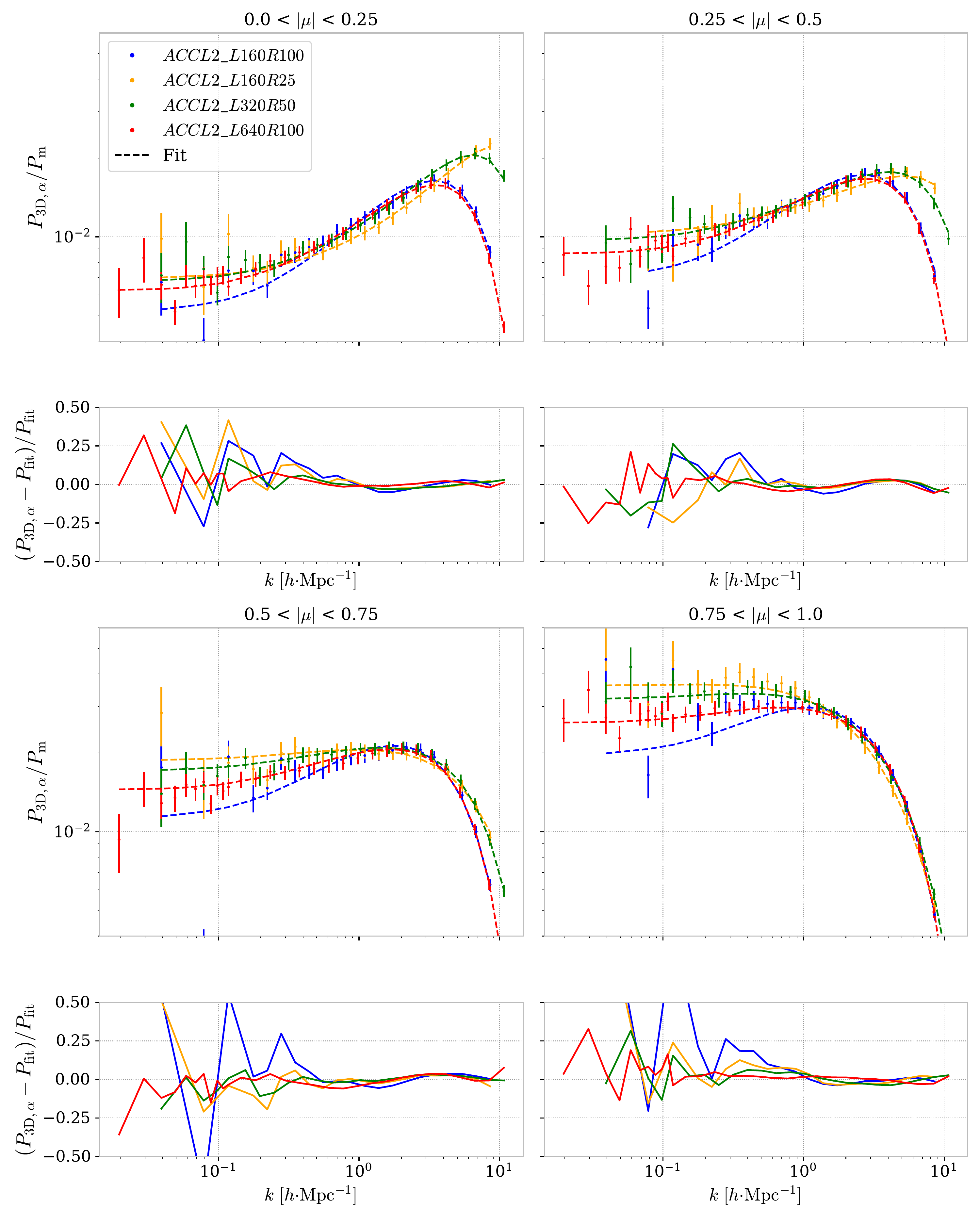}
    \caption{Simulated \pkt~(points) and their fit (dashed lines) of the $D_1$ non-linear parameterization along with their relative difference for the \textit{L160R100} (blue), \textit{L160R25} (yellow), \textit{L320R50} (green) and \textit{L640R100} (red) simulations in 4 $\mu$-bins at redshift $z=2.0$. All the power spectra are represented normalized by the linear matter power spectrum.}
    \label{fig:p3d_fit}
\end{figure*}

\begin{figure*}
    \centering
    \includegraphics[trim=0cm 0cm 0cm 0cm, clip=true,width = 1.0\textwidth]{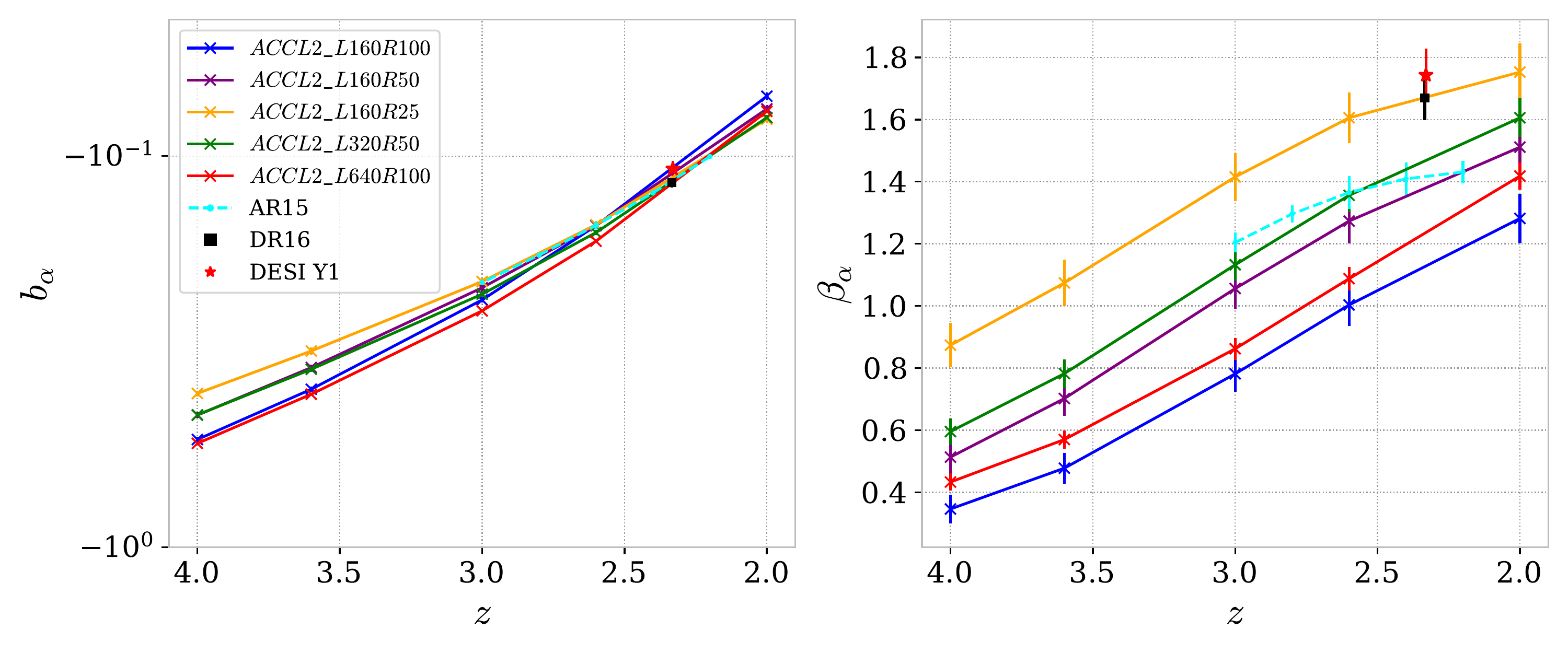}
    \caption{Redshift evolution for the linear parameters $b_{\alpha}$ (left panel) and $\beta_{\alpha}$ (right panel) for the \textit{L160R100} (blue), \textit{L160R50} (purple), \textit{L160R25} (yellow), \textit{L320R50} (green) and \textit{L640R100} (red) simulations when fitting their \pkt~with the $D_1$ non-linear parameterization compared with the same parameters measured in the theoretical work from~\citet{arinyo-i-prats_non-linear_2015} (light blue dashed line), those measured in eBOSS DR16 data~\citep{DuMas2020} (black square), and in DESI Y1 data~\citep{DESIlya2024} (red star).}
    \label{fig:linear_param_nosplice}
\end{figure*}

\begin{figure*}
    \centering
    \includegraphics[trim=0cm 0cm 0cm 0cm, clip=true,width = 1.0\textwidth]{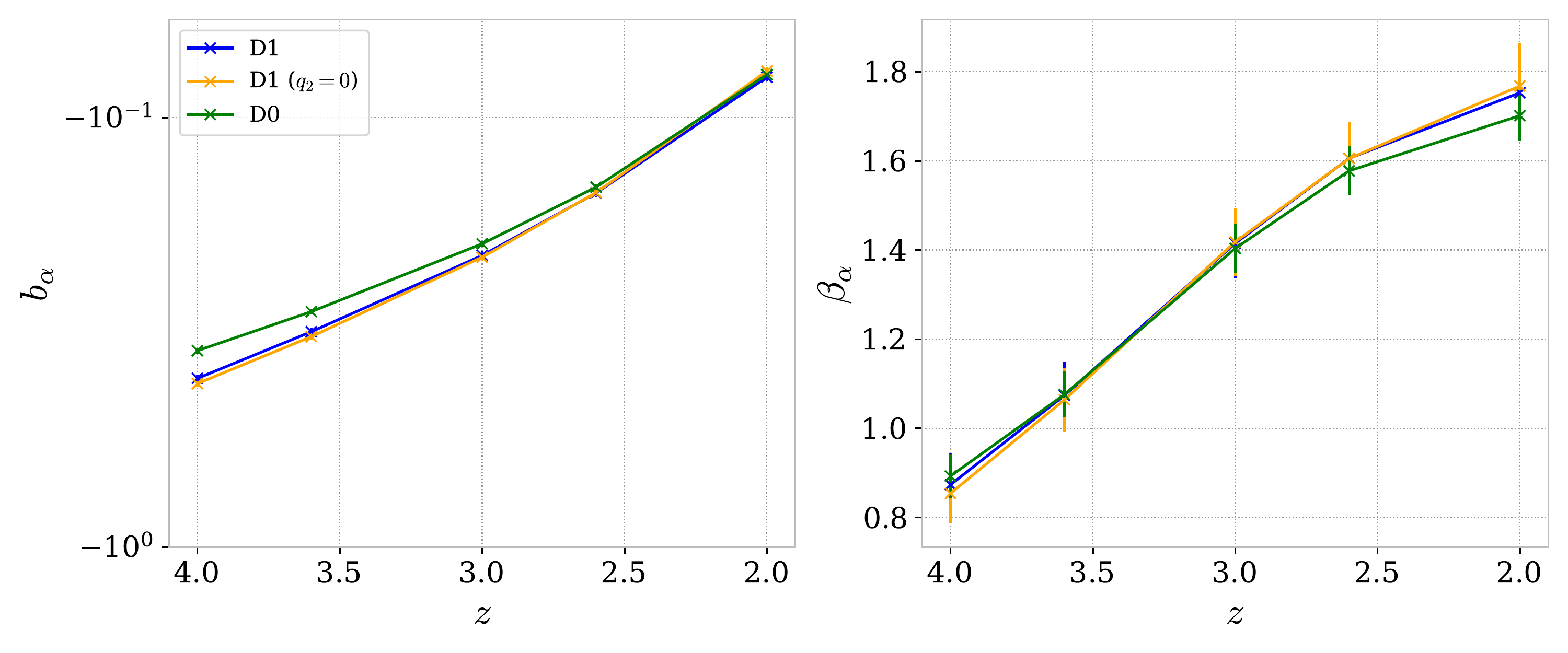}
    \caption{Redshift evolution for the linear parameters $b_{\alpha}$ (left panel) and $\beta_{\alpha}$ (right panel) for the \textit{L160R25} simulation when fitting its \pkt~with the $D_1$ (blue), $D_1(q_2=0)$ (yellow) and D0 (green) non-linear parameterization.}
    \label{fig:linear_param_nosplice_fit_variation}
\end{figure*}

\subsection{Fitting procedure}

In our study, we are interested in fitting medium and large scales of \pkt. In particular, we do not attempt to fit correctly the smallest scales in our simulations ($k \gtrsim 20$ \hpmpc) which are significantly influenced by the highly nonlinear physics of baryons. Similar to~\citet{arinyo-i-prats_non-linear_2015}, we apply several modifications to our \pkt~measurements before fitting the models outlined in Sec.~\ref{sec:models}.

\com{We first choose to redefine the range of wavenumber on which the fit will be realized. To reject the very-small scales, we first choose a maximal value of wavenumber for the fit $k_{\mathrm{c}} = 12$ \hpmpc, similar to~\citet{arinyo-i-prats_non-linear_2015}. As the wavenumber binning is linear, the intermediate scales ($1$ \hpmpc $\lesssim k < k_{\mathrm{c}})$ contains a lot more of wavenumber modes than large scales (approximately ten time more, varying on the considered grid)}. We apply a constant binning in $\log(k)$ to outweight the largest scales in the fit. For the smallest wavenumber, the \pkt~points are too sparse to impose this rebinning. Consequently, we choose to keep a linear binning for $k < k_{\mathrm{l}}$. We choose $k_{\mathrm{l}} = 0.12$ \hpmpc, and a number of 20 points in the rebinned part ($k>k_{\mathrm{l}}$), so that the number of bins is identical for $k>k_{\mathrm{l}}$ and $k<k_{\mathrm{l}}$.

Again to avoid giving excessive weight to the smallest scales, we follow the same prescription as in~\citet{arinyo-i-prats_non-linear_2015}: we modify the statistical uncertainty on \pkt:

\begin{equation}
    \label{eq:error}
    \sigma(k, \mu)=P_{3\mathrm{D},\alpha}(k, \mu)\left[\frac{1}{ \sqrt{N(k, \mu)}}+\epsilon\right]\,.
\end{equation}

\com{We take the same value as~\cite{arinyo-i-prats_non-linear_2015} such $\epsilon = 0.05$}. This additional term can be seen as an ad-hoc 5~\% systematic uncertainty on our capacity to model \pkt~at small scales. \com{Note that this uncertainty does not account for the cosmic variance associated with the specific seed we choose for the initial conditions of our simulations. The estimation of this contribution would require to run additional simulations with different seeds.}

We fit \pkt~by $\chi^2$ minimization using the \texttt{iminuit}~\citep{dembinski_2022} package, implemented in \texttt{lyapower} \git{corentinravoux/lyapower}{}, with a $\chi^2$ loss function defined by:

\begin{equation}
\label{eq:chi2}
\chi^{2}=\sum_{(k,\mu)} \frac{\left(P_{3\mathrm{D},\alpha}-P_{\mathrm{model},\alpha}\right)^{2}}{\sigma^{2}}\,,
\end{equation}

\noindent where for a given $(k,\mu)$ bin, the three-dimensional model $P_{\mathrm{model},\alpha}$ is expressed by integrating the model given in\com{ equations~\ref{eq:model_p3d},~\ref{eq:model_p3d_nl_0}, and~\ref{eq:model_p3d_nl_1}} over the whole $\mu$ range of the considered bin, and at the middle value of the k bin. We checked that integrating over $\mu$ is needed considering the large $\mu$ bins we are considering, and that integrating the model over the k bins does not change our fit.

Conversely to~\cite{arinyo-i-prats_non-linear_2015}, we directly use the error bars of the data $\sigma_{P}$ in the $\chi^2$ instead of the errors computed from the fitted model, to avoid any biasing during the fit.

\com{The $\chi^2$ function does not account for correlations between the measured $P_{3\mathrm{D},\alpha}$ in different $(k, \mu)$ bins. A proper numerical computation of those correlations, especially in the case of low-$k$ modes, would require running additional simulations with different seeds. As an alternative, one could estimate analytical or semi-analytical covariance of the three-dimensional power spectrum, but this is beyond the scope of this article. Finally, the addition of the $\epsilon$ parameter in Eq.~\ref{eq:error} introduces artificial correlations between the data points and we are not accounting for this in the $\chi^2$ function. As a consequence of all those considerations, the absolute value of the $\chi^{2}$, or the reduced $\chi^{2}$, can not be directly interpreted.}

The linear power spectrum $P_{\mathrm{m}}(k,z)$ used in the model is computed with the \texttt{CLASS} \com{software~\citep{class2011}}. Previous studies~\citep{arinyo-i-prats_non-linear_2015,givans_non-linearities_2022} used the matter power spectrum from the simulation itself at the initial redshift, and rescaled with a Boltzmann solver to the redshift of the fit. We tested on our smaller grids that using directly \texttt{CLASS} or the initial matter power spectrum from the simulation for the estimation of $P_{\mathrm{m}}$ does not change the fitted parameters. It is due to the high redshift our simulations are initialized ($z=200$), for which the linear matter power spectrum and the one computed on simulation are very similar. Furthermore, for the BAO study in Sec.~\ref{sec:bao}, we need a Boltzmann code estimation of the matter power spectrum, so for simplicity we chose to have the same estimation of $P_{\mathrm{m}}$ for both \pkt~modeling and BAO signal studies.

\subsection{Results}
\label{sec:fitresults}

Fig.~\ref{fig:p3d_fit} shows the simulated \pkt~and their fit with the $D_1$ non-linear parameterization along with their relative difference for the \textit{L160R100}, \textit{L160R25}, \textit{L320R50} and \textit{L640R100} simulations at $z=2.0$. As explained above, errors are larger at large scales because of cosmic variance, but smaller scales have more weight in the fit making the agreement between simulated \pkt~and fit much better at small scales, i.e. $k > 1$ \hpmpc. By eye, the fit looks equally good for all the simulations but we note a slightly better agreement along the lines of sight ($0.75 \leq |\mu| \leq 1.0$). An important finding is that \textit{L160R100} and \textit{L160R25} display significant differences ($\geq 15\%$) even at large scales whereas they have the same box size. This comes from the fact that small-scale structures impact the growth of large-scale modes through modes coupling with surprisingly more impact than the box size (e.g. differences are smaller between \textit{L640R100} and \textit{L160R100}).

Fig.~\ref{fig:linear_param_nosplice} shows the redshift evolution for the linear parameters $b_{\alpha}$ (left panel) and $\beta_{\alpha}$ (right panel) fitted from the \textit{L160R100}, \textit{L160R25}, \textit{L320R50} and \textit{L640R100} simulations compared to the same parameters measured from simulations of~\cite{arinyo-i-prats_non-linear_2015}, and to those measured from eBOSS DR16 data~\citep{DuMas2020}, and DESI DR1 data~\citep{DESIlya2024}. All fit results are provided in Tab.~\ref{tab:results_no_splice_1} and~\ref{tab:results_no_splice_2}. First, all simulations have the same redshift evolution, with $b_\alpha$ increasing with decreasing redshifts due mostly due to the mean flux evolution, imposed in our simulations by Eq.~\ref{eq:mean_flux}. At low redshifts, the bias is in very good agreement among the different simulations and with the one measured with data indicating that it does not depend on the box size or resolution. We note a 15-30\% difference at high redshifts between the different simulations, with the resolution being the most impactful parameter. Indeed, $b_{\alpha}$ is very similar for \textit{L160R100} and \textit{L640R100} but increases as we increase the resolution. As shown in~\cite{Chabanier2023}, resolution requirement for the \lya\ forest increases with redshift, as the signal comes from lower density region at high redshifts compare to lower redshifts. Finally, we note that while \textit{L160R100} and the simulation used in~\cite{arinyo-i-prats_non-linear_2015} are similar in terms of box size and resolution, the bias is a few percent higher for the \com{latter} and has a weaker redshift evolution.

On the right panel of Fig~\ref{fig:linear_param_nosplice}, we have much more variations in the fitted results of $\beta_{\alpha}$ but they all have a similar redshift evolution. The $\beta_{\alpha}$ term is sourced from peculiar velocities in the IGM by the RSD effect. The velocities get larger as the universe evolves and structures grow making gravitational attraction larger. Regarding dependence on the resolution and box size, a main result of this study is that the resolution has more impact on the values of linear parameters, in particular $\beta_\alpha$, than the box size. Indeed, $\beta_\alpha$ is $\sim$ 40\%  lower for \textit{L160R100} than \textit{L160R25} at all redshifts, while the differences between \textit{L160R100} and \textit{L640R100} are only about 5-10\%. Our high-resolution simulation \textit{L160R25} is in remarkable agreement with the BOSS and DESI measurements. The differences in the $\beta_\alpha$ are driven by the differences in the \pkt~at large scale modes that are strongly impacted by mode coupling. Therefore, we conclude that the convergence of the RSD parameter $\beta_\alpha$ is mainly driven by the physical resolution, more than the box size. It therefore appears that, as we increase resolution, we improve the characterization of structure growth at the smallest scales of our simulation, and this impacts the related linear bias. Therefore, an under-resolved simulation, like in~\cite{arinyo-i-prats_non-linear_2015}, does not properly account for peculiar velocities in the IGM, thus reducing $\beta_\alpha$. \com{However, we acknowledge that our simulations might not be totally converged in terms of resolution as shown in Fig.~\ref{fig:linear_param_nosplice}, especially at large redshift for which resolution requirements are stricter~\citep{bolton_resolving_2009,Chabanier2023,doughty_convergence_2023}. Finally, we note that the simulation in~\cite{arinyo-i-prats_non-linear_2015} have a less steep redshift evolution.}

We tested alternative fitting models developed in previous analysis~\citep{mcdonald_toward_2003,arinyo-i-prats_non-linear_2015} on our simulations. The details of the fitted parameters, as well as the $\chi^2$ and reduced $\chi^2$ values, are given in Tab.~\ref{tab:results_no_splice_fit_variation}. The $\chi^2$ is calculated according to equation \ref{eq:chi2}, and the reduced $\chi^2$ by dividing by the number of degrees of freedom. We note that the reduced $\chi^2$ values are mostly below one, which could be interpreted by over-fitting. \com{However, as discussed previously, those values cannot be easily interpreted due to the addition of the $\epsilon$ parameter in Eq.\ref{eq:error}, and the fact that we are neglecting correlations between error bars. In particular, increasing the $\epsilon$ parameter tends to increase error bars and largely decrease the $\chi^2$ values. For $\epsilon = 0$, the error bars of \pkt~are very close to zero at small scales due to the large number of wavenumber modes. In that configuration, the fit is totally driven by the small scales and gives disproportionate values of reduced $\chi^2$ (near 20). Furthermore, the error bars are underestimated at large scales as they do not account for cosmic variance. The latter effect increases the $\chi^2$ values. For all those reasons, we consider that we are not able to interpret the absolute level  of reduced $\chi^2$, and we only use those values for comparing different model performances.}

Fig.~\ref{fig:linear_param_nosplice_fit_variation} shows the $b_{\alpha}$ and $\beta_{\alpha}$ parameters for the \textit{L160R25} grid using our standard fitting procedure ($D_1$ non-linear term in ~\ref{eq:model_p3d_nl_1}) compared to $D_1(q_2=0)$ and $D_0$ (\ref{eq:model_p3d_nl_0}) parameterizations. The two $D_1$ models yield the same result for both the bias and the RSD parameter. The $D_0$ model is in disagreement for $b_{\alpha}$ only at large redshifts. We find that the fitted $D_0(k,\mu)$ profile does not reach unity even for the largest scales of our simulation ($k \lesssim 0.05$ \hpmpc), in accordance with findings from~\cite{arinyo-i-prats_non-linear_2015} and \cite{givans_non-linearities_2022}. It indicates that the $D_0$ model is not adequate for fitting, and thus can have an impact the value of bias.

In~\cite{arinyo-i-prats_non-linear_2015} and in~\cite{givans_non-linearities_2022}, the authors tend to prefer $D_1(q_2=0)$ model to reduce the degree of freedom of the fit. As shown in Tab.~\ref{tab:results_no_splice_fit_variation}, we see that both the $\chi^2$ and the reduced values are slightly lower for the $D_1$ values. We interpret this as the fact that we are fitting simulations with improved resolution compared to~\cite{arinyo-i-prats_non-linear_2015} and \cite{givans_non-linearities_2022}, and our \pkt~estimations should include more non-linear physics in our $k$ range because of modes coupling. The $\chi^2$ values for the $D_0$ model are lower than the others, but based on the previous consideration regarding the largest scales, we choose to discard this model. Furthermore, \com{some $D_0$ non-linear parameter values does not exhibit a regular variation as a function of redshift, i.e., they decrease and increase in a non consistent way}, which is not the case for the $D_1$ models. To conclude, we prefer to keep the $D_1$ model, considering that the $D_1(q_2=0)$ model is also viable. 

\section{Fits results on spliced power spectra}
\label{sec:rsdsplice}
\subsection{The splicing approach}
\label{sec:splice_method}

\begin{figure*}
    \centering
    \includegraphics[trim=0cm 0cm 0cm 0cm, clip=true,width = 0.75\textwidth]{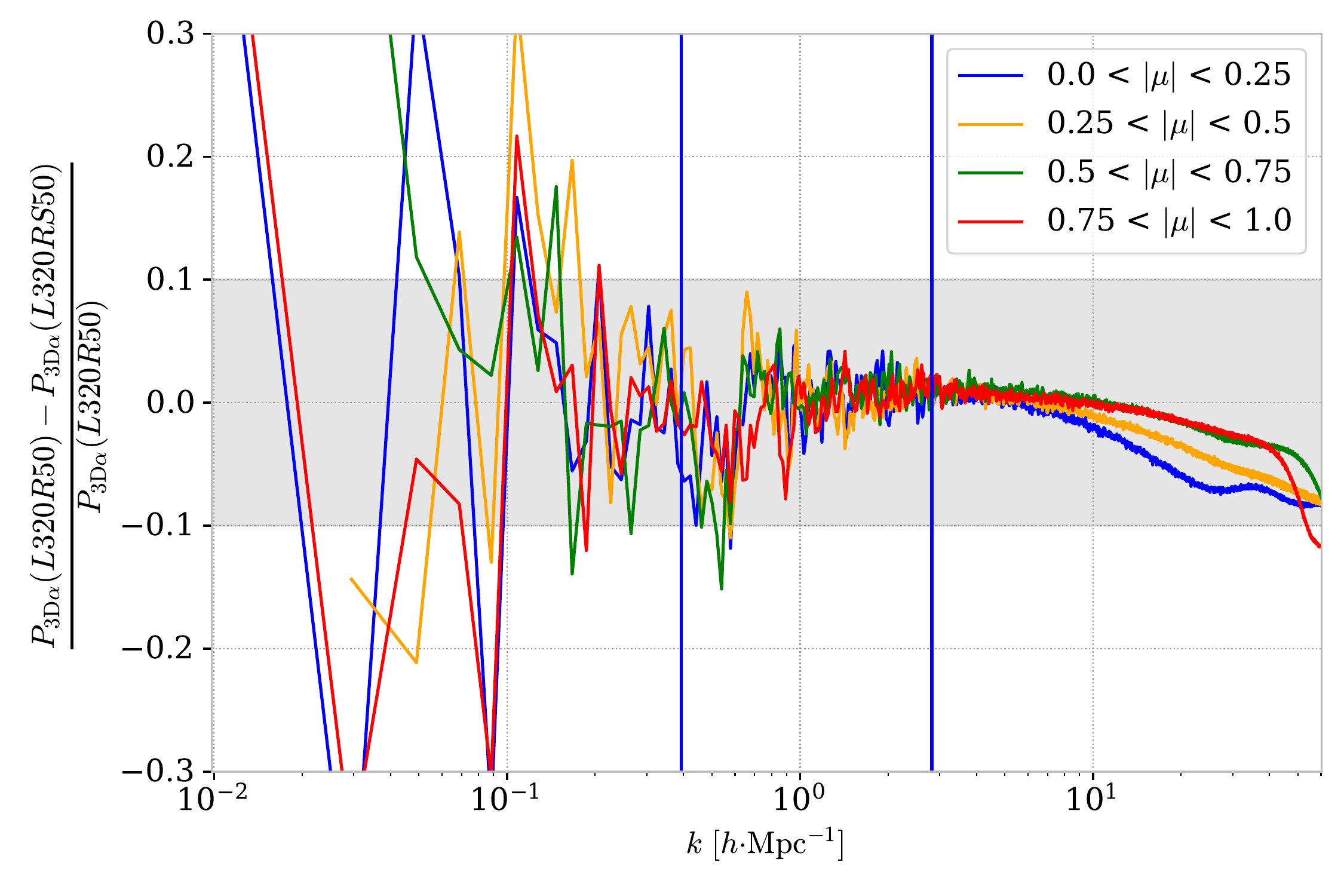}
    \caption{Absolute difference between the \pkt~measured from the true simulation (\textit{L320R50}) and the spliced simulation with identical resolution and size (\textit{L320SR50}), at redshift $z=2.0$. The gray band corresponds to a $10\%$ relative difference. The two blue vertical lines correspond to the parameters $k_{\mathrm{min}}$ (left) and $k_{\mathrm{max}}$ used for the splicing.}
    \label{fig:splice_verif}
\end{figure*}

\begin{figure*}
    \centering
    \includegraphics[trim=0cm 0cm 0cm 0cm, clip=true,width = 1.0\textwidth]{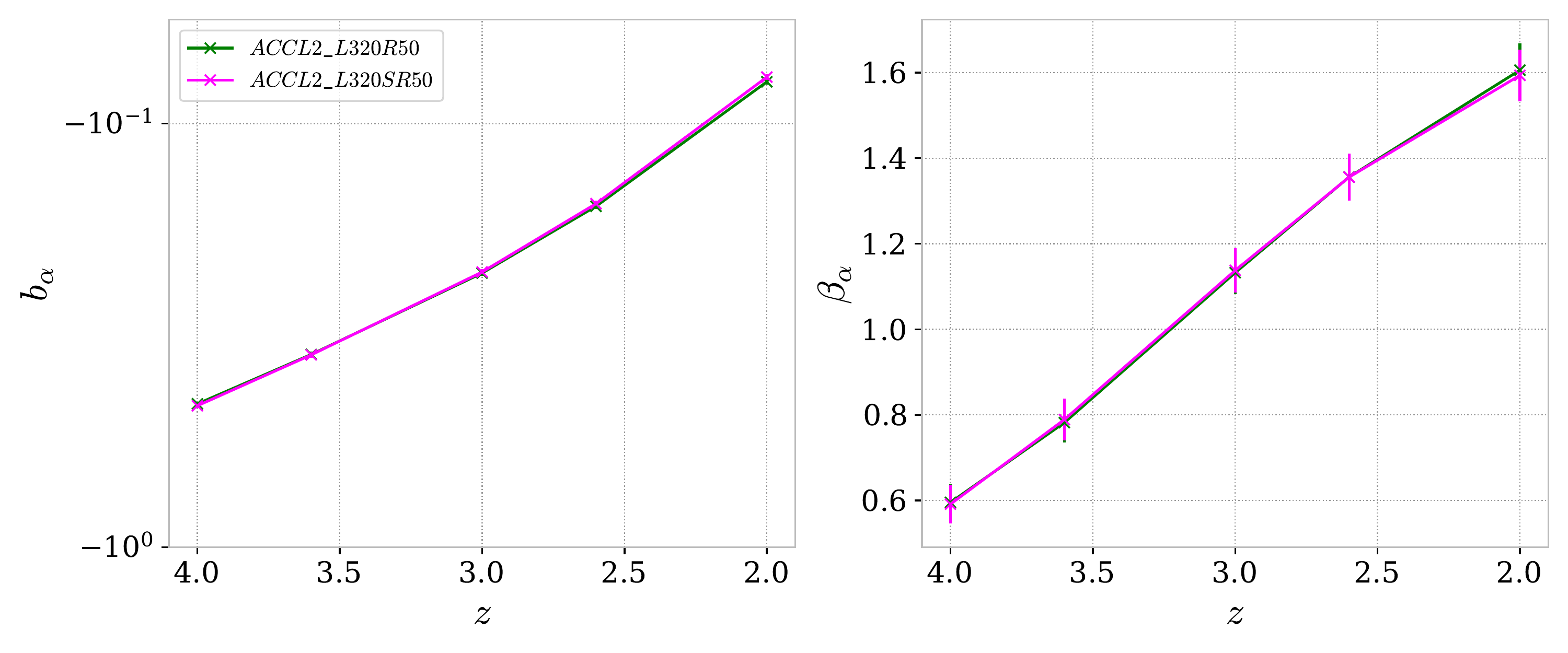}
    \caption{Redshift evolution for $b_{\alpha}$(left panel) and $\beta_{\alpha}$ (right panel) when fitting the true \pkt~from the \textit{L320R50} simulation (green) or its equivalent spliced \textit{L320SR50} (pink).}
    \label{fig:linear_param_splice_verification}
\end{figure*}

We wish to keep the small-scale information from \textit{L160R25} while using the large-scale information simulated by \textit{L640R100}. Splicing is a numerical trick that uses high-resolution simulations to correct the power spectrum of lower-resolution simulations with a larger size. This technique was used in~\citet{mcdonald_toward_2003} and~\citet{borde_new_2014} to estimate the three-dimensional and one-dimensional power spectrum, respectively.

Let us assume we have two simulations with large and small box sizes (respectively L and S) and low and high resolutions (respectively Lr and Hr). The splicing method consists of complementing the simulations (S,Hr) and (L,Lr) with a "cheap" simulation (S,Lr) to approximate the three-dimensional power spectrum of a hypothetical (L,Hr) simulation. This approximated power spectrum, so-called spliced, is defined using adapted rescalings for three different wavenumber regions, defined by limit values $k_{\mathrm{min}}$ and $k_{\mathrm{max}}$. The spliced power spectrum $P_{3\mathrm{D},\alpha}^{\mathrm{sp}}$ is expressed by

\begin{equation}
\label{eq:splicing}
P_{3\mathrm{D},\alpha}^{\mathrm{sp}} (k,\mu) = 
\begin{cases} P_{3\mathrm{D},\alpha}^{\mathrm{(L,Lr)}} (\Vec{k}) \dfrac{P_{3\mathrm{D},\alpha}^{\mathrm{(S,Hr)}} (k_{\mathrm{min}},\mu)}{P_{3\mathrm{D},\alpha}^{\mathrm{(S,Lr)}} (k_{\mathrm{min}},\mu)}, & \text{for } k < k_{\mathrm{min}} \,, \\[15pt] 
P_{3\mathrm{D},\alpha}^{\mathrm{(L,Lr)}} (\Vec{k}) \dfrac{P_{3\mathrm{D},\alpha}^{\mathrm{(S,Hr)}} (k,\mu)}{P_{3\mathrm{D},\alpha}^{\mathrm{(S,Lr)}} (k,\mu)}, & \text{for }  k_{\mathrm{min}} < k < k_{\mathrm{max}}\,, \\[15pt] 
P_{3\mathrm{D},\alpha}^{\mathrm{(S,Hr)}} (\Vec{k}) \dfrac{ P_{3\mathrm{D},\alpha}^{\mathrm{(L,Lr)}} (k_{\mathrm{max}},\mu)}{ P_{3\mathrm{D},\alpha}^{\mathrm{(S,Lr)}} (k_{\mathrm{max}},\mu)}, & \text{for } k > k_{\mathrm{max}}\,.
\end{cases}
\end{equation}

In the following, we note \textit{LXSRY}, results coming from the spliced $P_{3\mathrm{D},\alpha}^{\mathrm{sp}}$ of a simulation with box size $X$ \mpcph~and physical resolution artificially increased to $Y$ \kpcph~through the splicing technique.

\subsection{Splicing validation}
\label{sec:splice_verification}

We perform a first splicing to check the errors introduced by this method. We use a splicing that can be verified, i.e., for which we can also directly compute the \pkt~of a (L,Hr) simulation. The box sizes used are (L,S) $= (320,160)$ \mpcph~with resolutions (Hr,Lr) $= (50,100)$~\kpcph. The corresponding simulations are \textit{L320R100}, \textit{L160R50} and \textit{L160R100} so that we can compare the results from the truth \textit{L320R50} and its spliced equivalent \textit{L320SR50}.

We used this verification to optimize the value of the bounds $k_{\mathrm{min}}$ and $k_{\mathrm{max}}$. In~\citet{mcdonald_toward_2003}, those limit wavenumbers are defined by

\begin{equation}
\label{eq:k_splice}
\begin{aligned}
k_{\mathrm{min}} &= \frac{2\pi}{\mathrm{S}}\,,  \\
k_{\mathrm{max}} &= \frac{k_{\mathrm{Nyq}}}{4} = \frac{ \pi \mathrm{Lr} }{4\mathrm{L}}\,,
\end{aligned}
\end{equation}

\noindent where $k_{\mathrm{Nyq}}$ is the Nyquist frequency of the (L,Lr) simulation. Those definitions were valid for very small simulations with low resolution and are not adapted for the range of wavenumber of our study. Instead of the consideration used in~\citet{mcdonald_toward_2003}, we choose to define the $k_{\mathrm{min}}$ and $k_{\mathrm{max}}$ in a numerical way to reduce the error introduced by the splicing. A better recipe for $k_{\mathrm{max}}$ is to choose it as the last wavenumber for which the \pkt~difference between (L,Lr) and (S,Lr) simulations is lower than $1\%$. We verified that this new criteria minimizes the difference at small scales between the spliced and "truth" \pkt. In Eq.~\ref{eq:k_splice}, the initial definition of $k_{\mathrm{min}}$ implies that a very few numbers of wavenumbers are concerned by the cut $k < k_{\mathrm{min}}$, and that the computation of the splicing in this region is highly impacted by the large variations of \pkt~due to cosmic variance. By increasing this $k_{\mathrm{min}}$ value, we found that the difference between spliced and true \pkt~decreases. We choose a value $k_{\mathrm{min}} = 20\pi/\mathrm{S}$, which gives one of the lowest differences.

\begin{figure*}
    \centering
    \includegraphics[trim=0cm 0cm 0cm 0cm, clip=true,width = 1.0\textwidth]{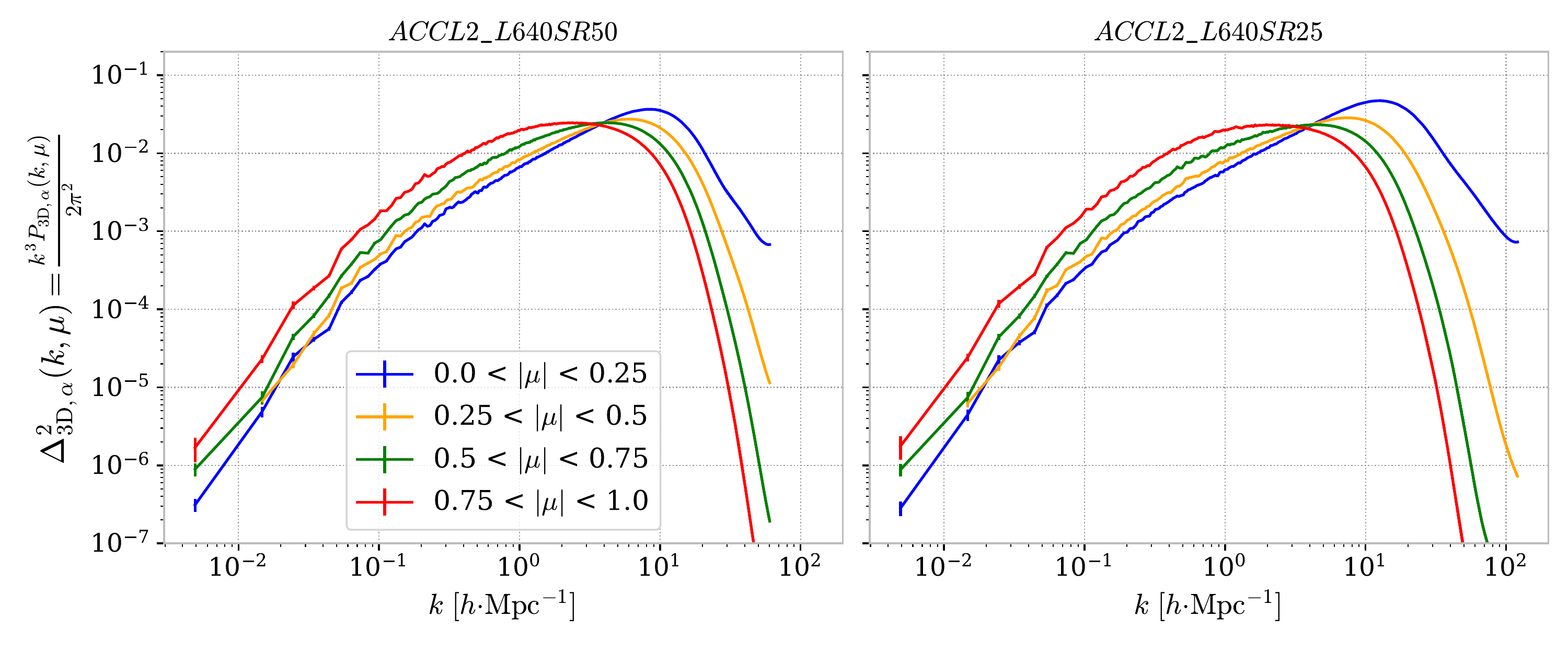}
    \caption{Dimensionless three-dimensional power spectrum ($\Delta_{3\mathrm{D}, \alpha}^2 = k^{3} P_{3\mathrm{D}, \alpha} / 2 \pi^2$) for the spliced power spectra \textit{L640SR50} (left) and \textit{L640SR25} (right). Both \pkt~are represented as a function of their wavenumber range (in \hpmpc) for different $\mu$ bins at redshift $z=2.0$.}
    \label{fig:p3d_plot_splice}
\end{figure*}

\begin{figure*}
    \centering
    \includegraphics[trim=0cm 0cm 0cm 0cm, clip=true,width = 1.0\textwidth]{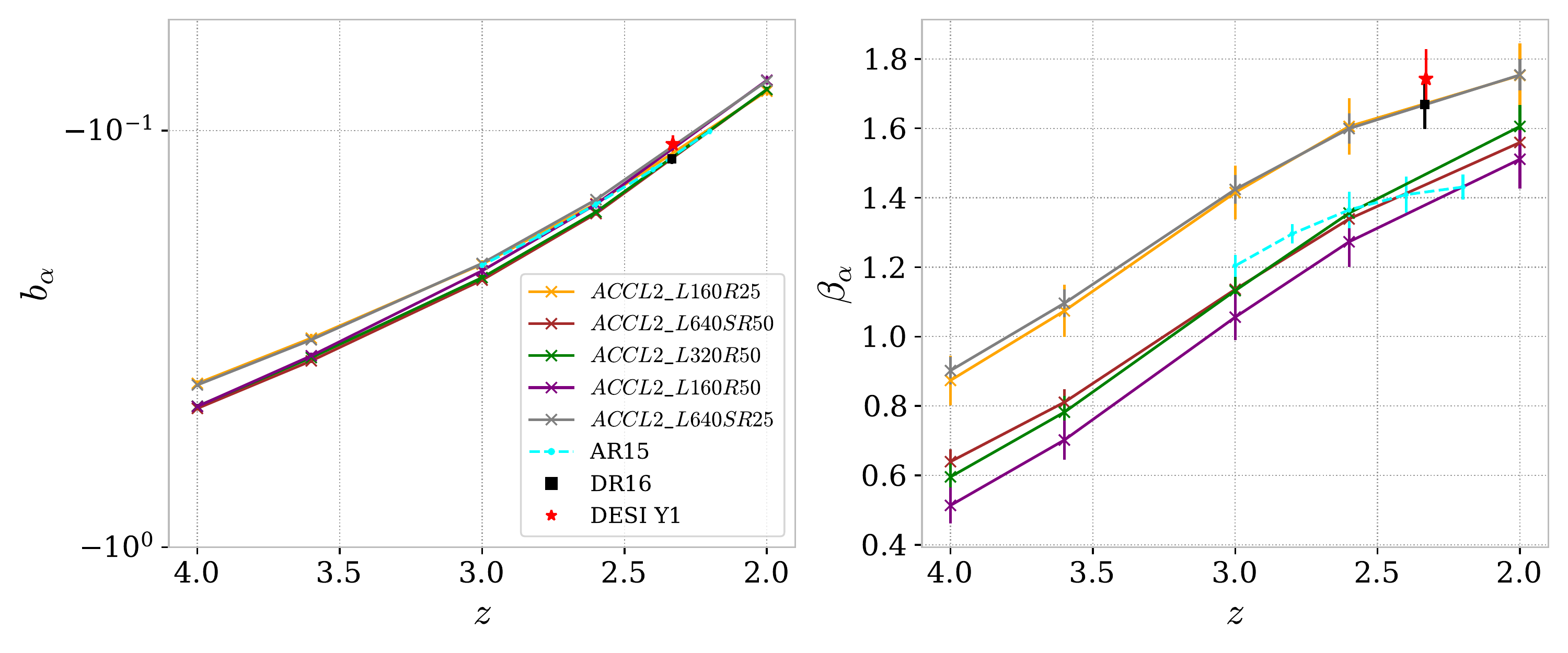}
    \caption{Redshift evolution for the linear parameters $b_{\alpha}$ (left panel) and $\beta_{\alpha}$ (right panel) from splicing \textit{L640SR50} (maroon) and \textit{L640SR25} (grey), compared to the results of the non-spliced simulations \textit{L160R25} (yellow) \textit{L320R50} (green), and \textit{L160R50} (purple) when fitting with the $D_1$ non-linear parameterization. We compared with the measurement from the theoretical work\com{ of~\citet{arinyo-i-prats_non-linear_2015} (light blue dashed line)}, from eBOSS DR16 data~\citep{DuMas2020} (black square), and in DESI Y1 data~\citep{DESIlya2024} (red star).}
    \label{fig:linear_param_splice}
\end{figure*}

Applying those updated wavenumber limits, the difference between the spliced \pkt~and the truth from the \textit{L320R50} output is shown in Fig.~\ref{fig:splice_verif}. For the smallest scales, $k\leq 10$ \hpmpc, the splicing method is precise at better than the 5\% level. Considering large scales, the difference is mainly dominated by the variations of the \pkt~estimation due to the low number of modes. Those variations are mainly due to the simulation's cosmic variance, which depends on the initial conditions chosen. Since the initial conditions are different between grid sizes $3072^3$ and $6144^3$, it is not possible to reproduce the large-scale variation of the \textit{L320R50} with the simulations we use for the splicing. Errors reach 20-30\% for the largest scales and are similar for all $\mu$ bins, but one should bear in mind that those scales have large error bars and have a low impact on the fit. More importantly, we compare in Fig~\ref{fig:linear_param_splice_verification} the fitted values of $b_\alpha$ and $\beta_\alpha$ when using the \pkt~ of the simulation or its splice at all redshifts. Both parameters are in remarkable agreement between the true \pkt~and the spliced one, indicating that we can safely use the splicing method to determine linear biases. The non-linear parameters, reported in Tab.~\ref{tab:results_fit_splice_verification}, present some discrepancy between the spliced and non-spliced versions, but those differences are negligible compared to the error bars of the fit. We note, however, that the $\chi^2$ values are better in the non-spliced case.

\subsection{Results}

We perform a second splicing for mimicking the \pkt~at the largest volume and intermediate resolution, i.e., $50$ \kpcph. We use box sizes (L,S) $= (640,320)$ \mpcph~with resolutions (Hr,Lr) $= (50,100)$~\kpcph. The corresponding simulations are \textit{L640R100}, \textit{L320R50} and \textit{L320R100}, and the spliced power spectrum is \textit{L640SR50}. We perform a third splicing using the higher-resolution simulation (\textit{L160R25}), our largest (\textit{L640R100}), and the simulation \textit{L160R100}. This splicing is called \textit{L640SR25}, and the resolution improvement is four times higher. We note that the splicing method was verified only for a factor two in resolution improvement, which must be considered when interpreting the \textit{L640SR25} splicing. The spliced power spectra of \textit{L640SR50} and \textit{L640SR25} are shown in Fig.~\ref{fig:p3d_plot_splice}, and the fitted parameters are given in Tab. \ref{tab:results_fit_splice}. Comparing those \pkt~with the Fig.~\ref{fig:p3d_plot}, it is clear that the splicing method allows to measure \pkt~over a wider dynamic range. The two spliced \pkt~exhibit similar features, but \textit{L640SR25} shows a cleaner separation between $\mu$ binning at small scales due to the improved resolution.

Fig~\ref{fig:linear_param_splice} shows the redshift evolution of the bias and RSD parameters as derived from the spliced \pkt~from \textit{L640SR50} along with the results from the non-spliced \pkt~from \textit{L160R25}, \textit{L320R50} and \textit{L160R50}, as well as the same values from the  \textit{L640SR25} splicing. For the three \pkt~with the same resolution (true or artificially increased for the spliced power spectra), as already observed in Sec.~\ref{sec:fitresults}, the \lya\ bias seems already converged for small boxes and at low resolution, so the biases measured with the spliced power spectra are almost exactly matching those from the other simulations. We note that the \textit{L640SR25} has a different bias value at high redshift, in agreement with the findings in Sec.~\ref{sec:fitresults}. The convergence of $\beta_\alpha$ is mainly driven by the physical resolution as the three same-resolution \pkt~have a $\beta_\alpha$ only a few percent different, i.e. $\sim 5$\% differences between \textit{L320R50} and \textit{L160R50}, and $\sim 2$\% differences between \textit{L640SR50} and \textit{L320R50}. The very small variation between the two latter \pkt~tends to indicate that we are converged at the percent level in terms of box size.

These results agree with our previous findings, i.e., that the resolution is the main driver of the convergence of the RSD parameter. This is striking when comparing the linear parameters for the \textit{L640SR25} and \textit{L160R25} which are in almost perfect agreement in Fig.~\ref{fig:linear_param_splice}. As splicing have a very small impact on the linear and non-linear parameters estimation, we consider the \textit{L160R25} as our best fitting results since it does not introduce potential errors that can be associated with a four-time resolution improvement in the splicing process. In conclusion, we refer the reader to Tab.~\ref{tab:results_no_splice_fit_variation} for the best estimation of parameters with the $D_1$ and $D_1(q_2=0)$ models.

\section{The BAO imprint on the 3D power spectrum}
\label{sec:bao}

\begin{figure*}
    \centering
    \includegraphics[trim=0cm 0cm 0cm 0cm, clip=true,width = \textwidth]{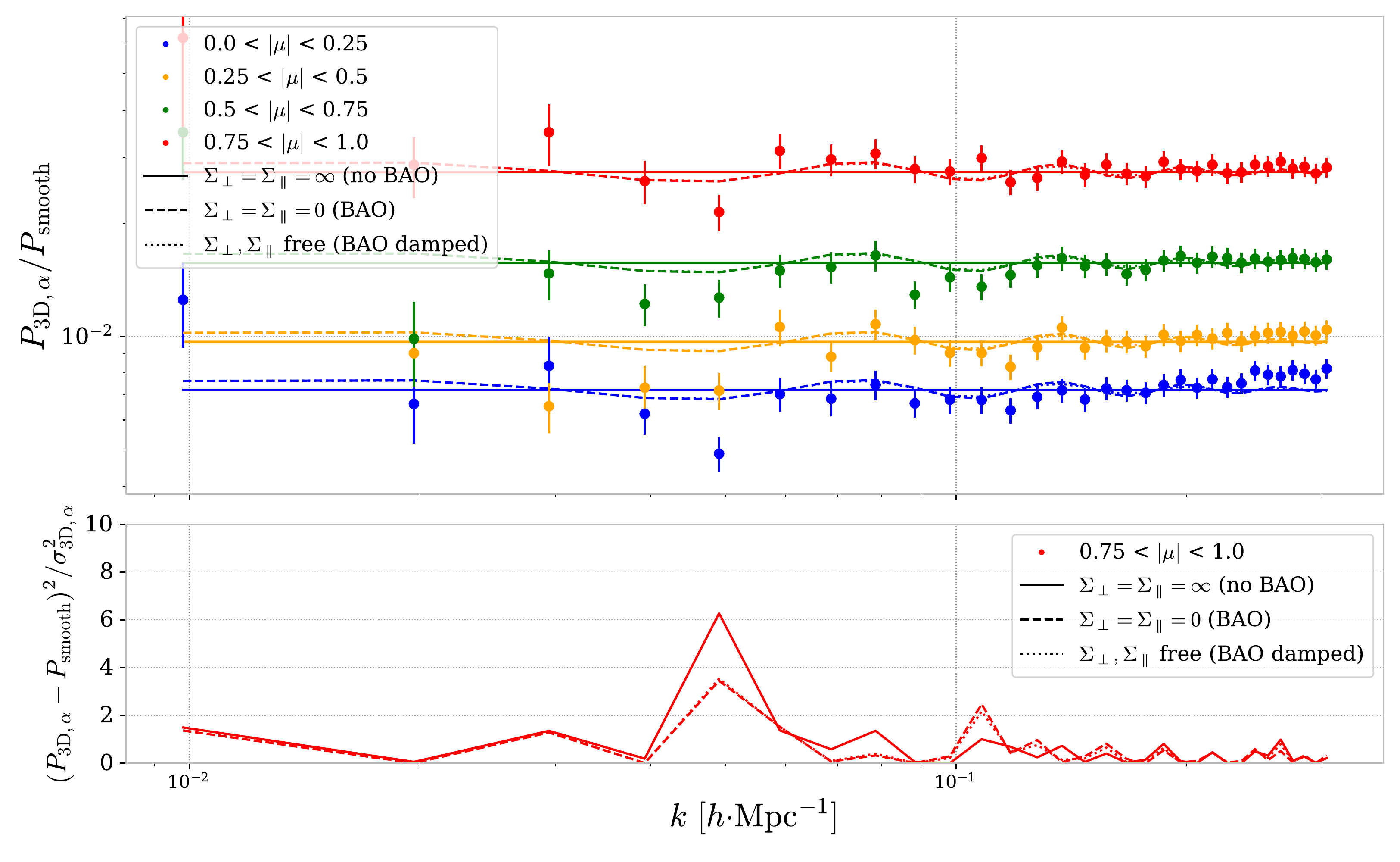}
    \caption{Measurements of the BAO feature from the \pkt of the \textit{L640R100} simulation at redshift $z=2.0$. Top: The points represents the ratios between \pkt and the model smoothed matter power spectrum $P_{\text{smooth}}$, for 4 $\mu$ bins. The fits associated with the $\Sigma_{\bot} = \Sigma_{\parallel} = \infty$ (no BAO), $\Sigma_{\bot} = \Sigma_{\parallel} = 0$ (BAO), and $\Sigma_{\bot}, \Sigma_{\parallel}$ free (BAO damped) models are respectively represented by full, dashed and dotted lines. \com{Note that the (BAO) and (BAO damped) represented by dashed and dotted lines are largely overlapping}. (Bottom) Residuals between the points and the models normalized by the error bars $\sigma(k,\mu)$ for the bin $ 0.75 < |\mu| < 1.0$. Only one $\mu$ bin is represented for clarity.}
    \label{fig:bao_fit}
\end{figure*}

\begin{figure*}
    \centering
    \includegraphics[trim=0cm 0cm 0cm 0cm, clip=true,width = 0.7\textwidth]{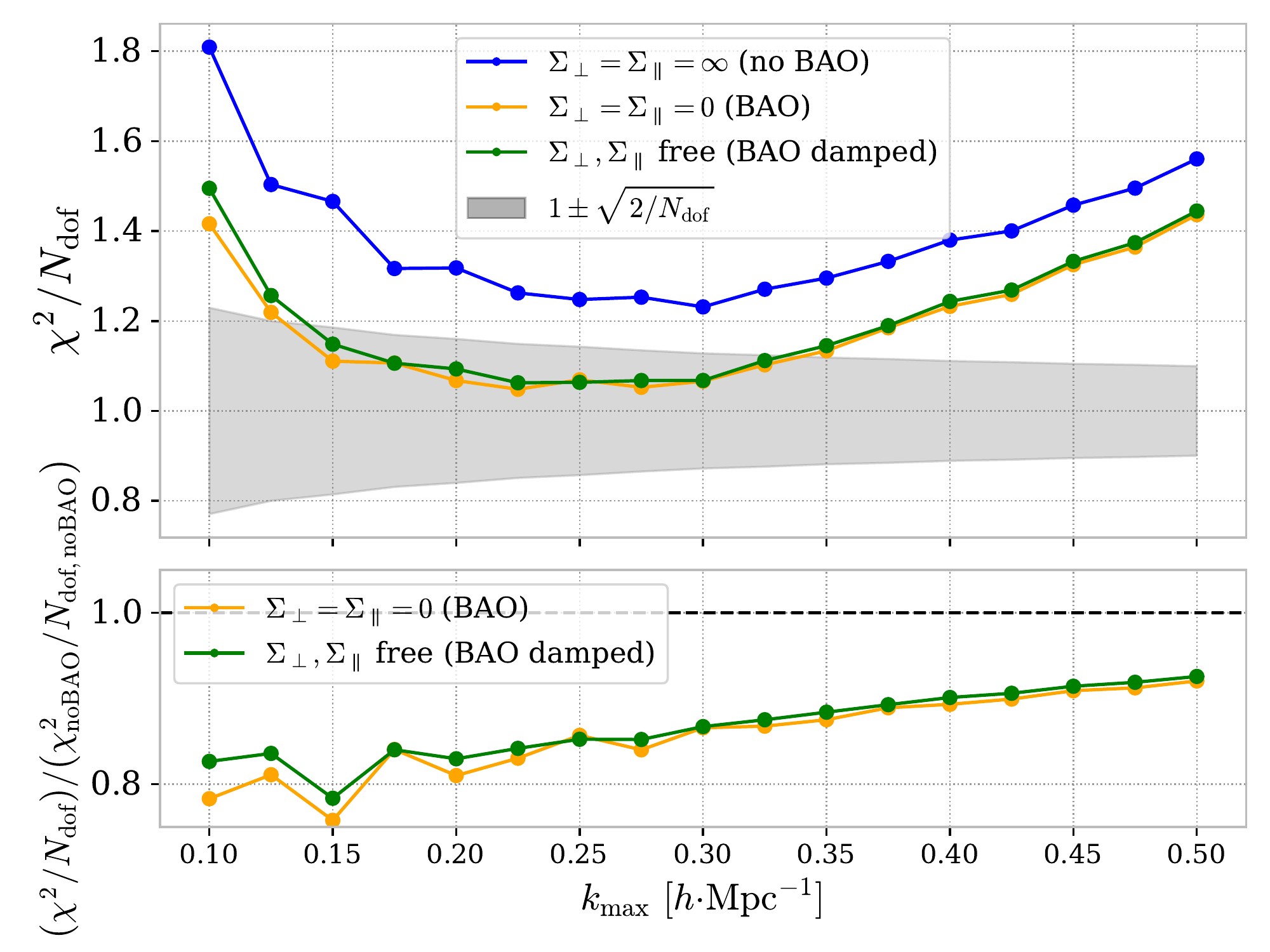}
    \caption{(Top) Reduced $\chi^2$ for the (no BAO), (BAO), and (BAO damped) models as a function of the maximal wavenumber considered in the fit $k_{\mathrm{max}}$, for the \textit{L640R100} simulation at redshift $z=2.0$. As an indication, the $1 \pm \sqrt{2 / N_{\mathrm{dof}}}$ region for the degree of freedom $N_{\mathrm{dof}}$ of (no BAO) and (BAO) models is shown as a shaded area. (Bottom) Ratios of reduced $\chi^2$ for (BAO) and (BAO damped) models over the one derived for the (no BAO) model.}
    \label{fig:bao_chi2}
\end{figure*}

\begin{figure*}
    \centering
    \includegraphics[trim=0cm 0cm 0cm 0cm, clip=true,width = 0.7\textwidth]{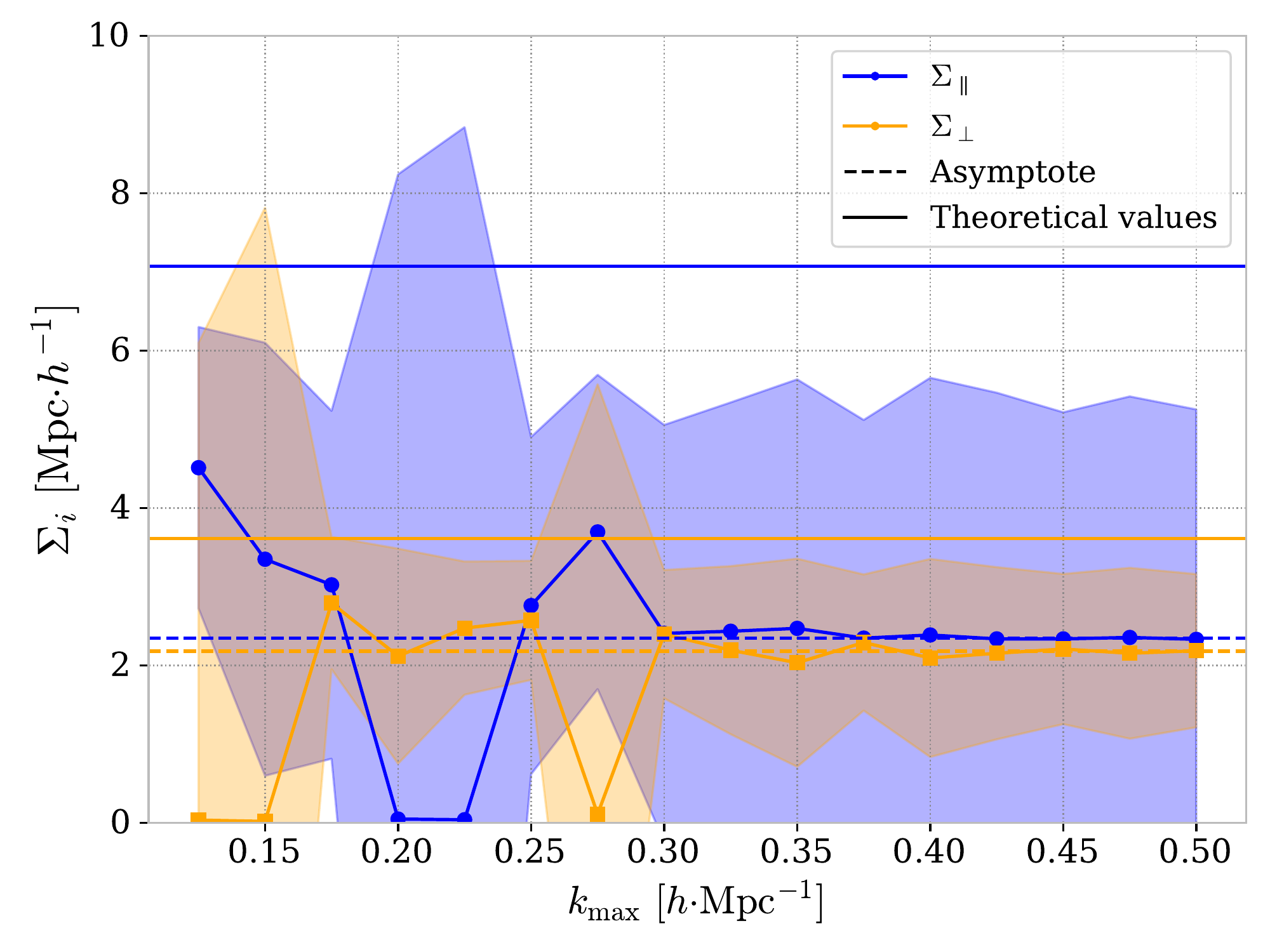}
    \caption{Measurements of the damping parameters $\Sigma_\perp$ and $\Sigma_\parallel$ as a function of $k_{\mathrm{max}}$, from the \textit{L640R100} simulation at redshift $z=2.0$. The associated 1$\sigma$ error bars are given by the \texttt{minuit} fitter. The asymptotic value of each parameter (here given by averaging for $k_{\mathrm{max}}$ larger than $0.3$ \hpmpc) is shown by a dashed line, and the theoretical values calculated from Eq.~\ref{eq:sigma_nl} and \texttt{CLASS}~\citep{class2011} are shown with full lines.}
    \label{fig:bao_damping}
\end{figure*}

Measuring three-dimensional correlations in the flux fluctuations of the \lya~forest provides an accurate method for measuring the scale of Baryon Acoustic Oscillations (BAO) signal. Given the scales reached by the simulations developed in this study, we want to measure the impact of the BAO on the \pkt. A goal of interest in the long run is to measure the smoothing induced by non-linear structure growth on the BAO signal. This smoothing was first estimated on N-body simulations in~\cite{Eisenstein2006}. It was first derived on \lya~forest data with BOSS~\citep{kirkby_fitting_2013} and added to all the modeling of large-scales \lya~forest correlations (see e.g.~\cite{bourboux_completed_2020} and~\cite{DESIlya2024} for more recent studies). From the results of our previous sections, we have seen that the non-linear clustering of the \lya~forest can significantly impact the large scales due to mode coupling. The statistical detection of BAO, both from survey data and simulation, requires very large comoving volumes: given their limited box size, previous hydrodynamical simulations dedicated to the \lya~forest could not achieve such a detection. The objective of this section is to measure the BAO signal along with its non-linear damping using our largest hydrodynamical simulation \textit{ACCL2$\_$L640R100}, and characterize potential deviation from N-body simulations~\citep{Eisenstein2006}.

\subsection{BAO signal modeling}

The models used in the previous sections do not enable us to suppress the BAO signal or to incorporate a BAO damping caused by non-linear growth. Therefore, we redefine the $P_m(k)$ term in the Kaiser formula (Eq.~\ref{eq:model_p3d}) to model the BAO signal, including its non-linear damping, and detect it in the \textit{ACCL2$\_$L640R100} simulation. As we focus on this section on the BAO signal only, we consider that the resolution of \textit{ACCL2$\_$L640R100} is good enough for this study.

To isolate and identify the BAO signal, a typical modeling method both in galaxies and \lya~forests~\citep{Eisenstein2006,kirkby_fitting_2013,eBOSS2020} consists of separating the matter power spectrum into a smooth component (noted $P_{\text{smooth}}$) and a part containing wiggles caused by BAO ($P_{\text{wiggles}}$). We performed this decomposition using the python package \texttt{cosmoprimo}\git{cosmodesi/cosmoprimo}{} which applies a power spectrum filter as implemented in \cite{Wallisch2018} and provides the two components mentioned above independently. The original matter power spectrum, taken before smoothing, is generated as in Sec.~\ref{sec:rsd}, i.e., using the \texttt{CLASS}~\citep{class2011}. We then incorporate the non-linear damping of the BAO signal with a Gaussian function, such that the matter power spectrum in Eq.~\ref{eq:model_p3d} becomes:

\begin{equation}
\label{eq:damping_power}
P_m(k,\mu) = P_{\text{smooth}}(k) + P_{\text{wiggles}}(k) \times \exp \left(-\dfrac{k^2 \Sigma_{\text{nl}}^2(\mu)}{2}\right)
\end{equation}

\noindent where $\Sigma_{\text{nl}}^2 = \Sigma_\parallel^2\mu^2 + \Sigma_\perp^2 (1- \mu^2) $. The smoothing effects of non-linear bulk motions on the BAO can be described by the parameters $\Sigma_\parallel$ along the line-of-sight and $\Sigma_\perp$ transverse to the line-of-sight. These damping terms result in a decrease in the amplitude of the BAO signal oscillations present in the linear power spectrum template $P_{\text{wiggles}}$, making the BAO feature less prominent and more challenging to detect. They are directly caused by non-linear structure formation at the redshift we are considering and were extensively described for galaxies \com{\citep{Eisenstein2006,Seo2007}, and \lya~forest~\citep{kirkby_fitting_2013}}.

As we are focusing on the largest scales of \pkt ($k \leq 0.5$ \hpmpc), we do not take into account the non-linear term $D(k,\mu)$ in Eq.~\ref{eq:model_p3d}, and use the BAO decomposition Eq~\ref{eq:damping_power} with the linear part of the \pkt~model (the Kaiser formula). This model noted (BAO damped), can be turned into a model without BAO oscillations, noted (no BAO) by setting the damping parameters to a very high value $\Sigma_\parallel = \Sigma_\perp = \infty$, or a model without non-linear damping, noted (BAO) by settings $\Sigma_\parallel = \Sigma_\perp = 0$. For verification purposes, we also performed the study of this section with the non-linear term $D(k,\mu)$. We found that the results were independent of this term and that its inclusion induced significant instability in the fit.

\subsection{Detection of the BAO signal}

We performed fits of the (no BAO), (BAO), and (BAO damped) models, using \pkt resulting from the \textit{L640R100} simulation at redshift $z=2.0$. We are only interested in the largest scales for which the BAO signal is present, so we are focusing on a smaller wavenumber range than in the previous section. We vary the maximal wavenumber used in the fits, $k_{\mathrm{max}}$, from $0.1$ to $0.5$ \hpmpc. This is the typical wavenumber range where the BAO signal in the matter power spectrum is present. Contrarily to previous sections, we do not perform a constant $\log(k)$ re-binning, as it can suppress the BAO signal at smaller scales. 

The result of the fit with $k_{\mathrm{max}} = 0.3$ \hpmpc~ on the four $\mu$ bins for the three models are shown in Fig.~\ref{fig:bao_fit}. For the data points and the models, we divide \pkt~by $P_{\text{smooth}}$ to highlight the oscillations of the BAO signal at the expected range around $k \sim 0.1$ \hpmpc. The fit can capture the visible BAO signal for most considered scales. We note a discrepancy at the smaller scale considered for transverse correlations ($\mu \sim 0$) due to the appearance of non-linearities that the current model is not accounting for. The difference between (BAO) and (BAO damped) models is small and can only be seen at high wavenumbers.

In order to better quantify the detection significance of the BAO signal, we compute the reduced $\chi^2$ values for each model and different $k_{\mathrm{max}}$ values considered. They are shown in Fig.~\ref{fig:bao_chi2}, together with the reduced $\chi^2$ ratio between the models for which BAO is considered and the (no BAO) model. Over all the $k_{\mathrm{max}}$ range, we clearly see that the data points prefer a model with BAO oscillation than without. The (BAO) and (BAO damped) models then exhibit reduced $\chi^2$ values close to $1$ when fitting the \pkt with $0.15 \leq k_{\mathrm{max}} \leq 0.3$. As highlighted in Fig.~\ref{fig:bao_chi2}, for this $k_{\mathrm{max}}$ interval, the reduced $\chi^2$ profile of (BAO) and (BAO damped) models are within the $1 \pm \sqrt{2/N_{\mathrm{dof}}}$ range, where $N_{\mathrm{dof}}$ is the number of degree of freedom for the (BAO) model. This suggests that a "good" fit is obtained when including the BAO feature in the model, while the quality of the fit is less satisfying without a BAO feature. For higher $k_{\mathrm{max}}$ values, the goodness of fit of the models is degraded because we do not include the non-linear terms in this study. 

To assess the significance of the BAO detection, we compute the $\chi^2$ differences $\Delta \chi^2$ between the (BAO damped) and (no BAO) models. The $\chi^2$ statistics with the results of the fit for the two considered models results in a p-value $p < 0.0006$ for all the $k_{\mathrm{max}}$ considered here, inducing a clear detection of the BAO signal.

\subsection{Non-linear damping parameters estimation}

Fig.~\ref{fig:bao_damping} shows the fitted damping parameters $\Sigma_\parallel$ and $\Sigma_\perp$, and their associated 1$\sigma$ error range, as a function of $k_{\mathrm{max}}$ for the (BAO damped) model. The measured values are stable as a function of $k_{\mathrm{max}}$ for $k_{\mathrm{max}} \geq 0.3$ \hpmpc. We interpret this as the fact that the damping is mainly impacting the smaller scales of the BAO signal due to the exponential term in the Eq.~\ref{eq:damping_power}, so that the fitter can only correctly measure the damping for a large enough $k_{\mathrm{max}}$ value. Therefore, we considered the $k_{\mathrm{max}}$-asymptotic values to be our study's measured damping term values.

Following previous studies~\citep{Eisenstein2006,Seo2007,kirkby_fitting_2013}, we can theoretically estimate the damping parameters as

\begin{align}
    \label{eq:sigma_nl}
    &\Sigma_\perp = 10.4 D(z) \sigma_8 \\
    &\Sigma_\parallel = (1 + f) \Sigma_\perp\,,
\end{align}

\noindent where $D(z)$ represents the linear growth factor, and $f$ is the logarithmic growth rate. The values computed using \texttt{CLASS}~\citep{class2011} for the cosmological model of our simulation are reported in Fig.~\ref{fig:bao_damping}. Our measured values are in-between the theoretical ones and the case of no damping, but with large uncertainties. Quantitatively, considering the asymptotic values and error bars, we report a detection of $\Sigma_\perp$ (resp. $\Sigma_\parallel$) above zero at the $2.2\sigma$ (resp. $0.8\sigma$) level. Our fitted values are in agreement with the theoretical predictions from Eq.~\ref{eq:sigma_nl} at the $1.4\sigma$ (resp. $1.6\sigma$) level for $\Sigma_\perp$ (resp. $\Sigma_\parallel$). We conclude that, due to the large statistical uncertainty of our \pkt~measurement in the wavenumber range of $k$ that is relevant for BAO, we only have a mild hint of detection for the non-linear BAO damping. However, we provide neither a clear detection of this damping term, nor a tension with the theoretical prediction from Eq.~\ref{eq:sigma_nl}.

\section{Conclusion}
\label{sec:conclusion}
In this work, we have presented the most precise calculation of the three-dimensional \lya\ power spectra over a range of scales exceeding three orders of magnitude, using one of the largest simulations of the IGM to date, as well as one with a very high physical resolution. The suite of \texttt{Nyx} cosmological hydrodynamical simulations, named \textit{ACCEL$^2$} simulations, comprises six simulations with box sizes ranging from 160 to 640 \mpcph\ and physical resolution ranging from 25 to 100 \kpcph.

We use \com{two common analytical models \com{studied in~\cite{mcdonald_toward_2003},~\cite{arinyo-i-prats_non-linear_2015}, and~\cite{givans_non-linearities_2022}}} that correct the linear Kaiser equation to include an empirical model for non-linear growth of structures, Jeans pressure smoothing and thermal broadening. We show these non-linear models can fit the measured power spectra from simulations from $z=2.0$ to $z=4.0$.

We study the convergence of the linear parameters, the \lya~bias $b_\alpha$ and the \lya~RSD parameter $\beta_\alpha$, with box size and physical resolution. While the bias only mildly depends on the simulation box size and resolution, at least for redshifts $z\leq3$, the RSD parameter is strongly impacted. We find that the anisotropy of the power spectrum at large scales, which determines $\beta_\alpha$, is dominantly driven by the physical resolution through mode coupling, more than by the box size itself. This result constitutes a major finding of our work. We show in Sec.~\ref{sec:fitresults} that at fixed physical resolution, varying the box size from 160 to 640 \mpcph~only impacts $\beta_\alpha$ by 5-10\%. On the other hand, at fixed box size, varying the physical resolution from 100 to 25 \kpcph~impacts $\beta_\alpha$ by 60\%. The $\beta_\alpha$ values obtained from our best resolution run are in better agreement with SDSS data ($\beta_\alpha = 1.669$ at $z=2.334$~\citep{DuMas2020}) and DESI DR1 data ($\beta_\alpha = 1.743$ at $z=2.33$~\citep{DESIlya2024}), in particular compared to results from previous simulations which may have significantly under-estimated this parameter due to the of lack of resolution. However, we would like to stress that we did not demonstrate convergence for our calculation of $\beta_\alpha$, which means that using even better resolution would potentially yield higher values for $\beta_\alpha$.

We also perform a splicing of our simulation set to compute a spliced \pkt~which benefits both from the largest scales reached and the highest resolution of our simulations. We show that the spliced \pkt~provides very similar fitted values for both linear and non-linear parameters. We concluded that our highest resolution simulation gives our best-estimated model parameters, which are provided in Tab.~\ref{tab:results_no_splice_fit_variation} for the $D_1$ and $D_1(q_2=0)$ models.

Finally, for the first time to our knowledge with a fully hydrodynamical simulation, we measured the impact of the BAO feature on \pkt. We report a clear detection of the BAO signal, and obtain a marginal detection of its damping. However, the large statistical error bars, \com{stemming} from the $\lesssim 1$~Gpc$^3$ box size, prevent us from quantitatively comparing this damping with predictions from theory~\citep{Eisenstein2006}.

The \textit{ACCEL$^2$} hydrodynamical simulations are timely as they give us the opportunity to test and improve three-dimensional \lya~power spectrum models, for which some measurements have been recently reported~\citep{AbdulKarim2023,Belsunce2024}, and is expected to be largely improved with data from DESI and other similar surveys.

This work clearly points toward roads of improvement. First, we need to establish confidence in measuring the RSD parameters from simulations by reaching its full convergence with respect to physical resolution, i.e., when there is no variation of $\beta_\alpha$ when increasing the resolution. Given that the box size effect is sub-dominant compared to the impact of resolution, this can be done using smaller box sizes, allowing us to push the resolution to very small scales ($\sim$ 10 \kpcph). Secondly, large scales are also significantly impacted by the early-time statistical description of density fluctuations of the different components. In this work, both dark matter and baryons have the same transfer functions for initial conditions which is expected to induce numerical errors~\citep{Hahn2021}. Improved initial conditions schemes should be used in future work.

One of the goals of this study is to create a numerical laboratory to test and improve the models employed to describe the \lya~forest correlations over an extensive range of scales. In particular, we think this simulation set provides an opportunity to improve the modeling of auto- and cross-correlations~\citep{DuMas2020,DESIlya2024} or assess promising theoretical models. As examples, recent studies such as standard non-linear perturbation theory~\citep{chen_lyalpha_2021} or effective field theory such as EFTofLSS~\citep{Ivanov2023} could be tested with our simulations in the context of the \lya~forest.

\section*{Acknowledgements}

We thank Patrick McDonald, Nathalie Palanque-Delabrouille, and Christophe Yèche for fruitful discussions and comments that lead to the improvement of this analysis. 

Main simulations used resources of the Oak Ridge Leadership Computing Facility at the Oak Ridge National Laboratory, which is supported by the Office of Science of the U.S. Department of Energy under Contract No. DE-AC05-00OR22725.
We also thank the National Energy Research Scientific Computing Center (NERSC) for providing us resources for running auxiliary simulations and analysis tasks, a U.S. Department of Energy Office of Science User Facility located at Lawrence Berkeley National Laboratory, operated under Contract No. DE-AC02-05CH11231.
We further acknowledge the "Institut du développement et des resources en informatique scientifique" (IDRIS) for granting us the opportunity to use the full GPU extension of the Jean-Zay supercomputer\footnote{\url{http://www.idris.fr/jean-zay/}} with a "Grands Challenge" project allowing us to run preparatory simulations for this study. 

This work was partially supported by the DOE’s Office of Advanced Scientific Computing Research and Office of High Energy Physics through the Scientific Discovery through Advanced Computing (SciDAC) program. The development of Nyx as an AMReX application was supported by the U.S. Department of Energy, Office of Science, Office of Advanced Scientific Computing Research, Applied Mathematics program under contract number DE-AC02005CH11231, and by the Exascale Computing Project (17-SC-20-SC), a collaborative effort of the U.S. Department of Energy Office of Science and the National Nuclear Security Administration.  

The authors also acknowledge support from grant ANR-16-CE31-0021. The project leading to this publication has received funding from Excellence Initiative of Aix-Marseille University - A*MIDEX, a French “Investissements d’Avenir” programme (AMX-20-CE-02 - DARKUNI).

\section*{Data Availability}

The simulation outputs presented in this paper will be made available upon reasonable request.



\bibliographystyle{mnras}
\bibliography{biblio} 

\begin{thebibliography}{}
\makeatletter
\relax
\def\mn@urlcharsother{\let\do\@makeother \do\$\do\&\do\#\do\^\do\_\do\%\do\~}
\def\mn@doi{\begingroup\mn@urlcharsother \@ifnextchar [ {\mn@doi@}
  {\mn@doi@[]}}
\def\mn@doi@[#1]#2{\def\@tempa{#1}\ifx\@tempa\@empty \href
  {http://dx.doi.org/#2} {doi:#2}\else \href {http://dx.doi.org/#2} {#1}\fi
  \endgroup}
\def\mn@eprint#1#2{\mn@eprint@#1:#2::\@nil}
\def\mn@eprint@arXiv#1{\href {http://arxiv.org/abs/#1} {{\tt arXiv:#1}}}
\def\mn@eprint@dblp#1{\href {http://dblp.uni-trier.de/rec/bibtex/#1.xml}
  {dblp:#1}}
\def\mn@eprint@#1:#2:#3:#4\@nil{\def\@tempa {#1}\def\@tempb {#2}\def\@tempc
  {#3}\ifx \@tempc \@empty \let \@tempc \@tempb \let \@tempb \@tempa \fi \ifx
  \@tempb \@empty \def\@tempb {arXiv}\fi \@ifundefined
  {mn@eprint@\@tempb}{\@tempb:\@tempc}{\expandafter \expandafter \csname
  mn@eprint@\@tempb\endcsname \expandafter{\@tempc}}}

\bibitem[\protect\citeauthoryear{Abareshi et~al.}{Abareshi
  et~al.}{2022}]{DESI2022}
Abareshi B.,  et~al., 2022, \mn@doi [Astron. J.] {10.3847/1538-3881/ac882b},
  164, 207

\bibitem[\protect\citeauthoryear{Abdul-Karim, Armengaud, Mention, Chabanier,
  Ravoux  \& Luki\'c}{Abdul-Karim et~al.}{2024}]{AbdulKarim2023}
Abdul-Karim M.~L.,  Armengaud E.,  Mention G.,  Chabanier S.,  Ravoux C.,
  Luki\'c Z.,  2024, \mn@doi [JCAP] {10.1088/1475-7516/2024/05/088}, 05, 088

\bibitem[\protect\citeauthoryear{{Alcock} \& {Paczynski}}{{Alcock} \&
  {Paczynski}}{1979}]{Alcock1979}
{Alcock} C.,  {Paczynski} B.,  1979, \mn@doi [\nat] {10.1038/281358a0}, \href
  {https://ui.adsabs.harvard.edu/abs/1979Natur.281..358A} {281, 358}

\bibitem[\protect\citeauthoryear{{Almgren}, {Bell}, {Lijewski}, {Luki{\'c}}  \&
  {Van Andel}}{{Almgren} et~al.}{2013}]{Almgren2013}
{Almgren} A.~S.,  {Bell} J.~B.,  {Lijewski} M.~J.,  {Luki{\'c}} Z.,   {Van
  Andel} E.,  2013, \mn@doi [\apj] {10.1088/0004-637X/765/1/39}, \href
  {https://ui.adsabs.harvard.edu/abs/2013ApJ...765...39A} {765, 39}

\bibitem[\protect\citeauthoryear{Arinyo-i Prats, Miralda-Escud{\'e}, Viel  \&
  Cen}{Arinyo-i Prats et~al.}{2015}]{arinyo-i-prats_non-linear_2015}
Arinyo-i Prats A.,  Miralda-Escud{\'e} J.,  Viel M.,   Cen R.,  2015, \mn@doi
  [Journal of Cosmology and Astroparticle Physics]
  {10.1088/1475-7516/2015/12/017}, 2015, 017

\bibitem[\protect\citeauthoryear{Armengaud, Palanque-Delabrouille, Y{\`e}che,
  Marsh  \& Baur}{Armengaud et~al.}{2017}]{armengaud_constraining_2017}
Armengaud E.,  Palanque-Delabrouille N.,  Y{\`e}che C.,  Marsh D. J.~E.,   Baur
  J.,  2017, \mn@doi [Monthly Notices of the Royal Astronomical Society]
  {10.1093/mnras/stx1870}, 471, 4606

\bibitem[\protect\citeauthoryear{Baur, Palanque-Delabrouille, Y{\`e}che,
  Boyarsky, Ruchayskiy, Armengaud  \& Lesgourgues}{Baur
  et~al.}{2017}]{baur_constraints_2017}
Baur J.,  Palanque-Delabrouille N.,  Y{\`e}che C.,  Boyarsky A.,  Ruchayskiy
  O.,  Armengaud {\'E}.,   Lesgourgues J.,  2017, \mn@doi [Journal of Cosmology
  and Astroparticle Physics] {10.1088/1475-7516/2017/12/013}, 2017, 013

\bibitem[\protect\citeauthoryear{Bautista et~al.,}{Bautista
  et~al.}{2017}]{Bautista2017}
Bautista J.~E.,  et~al., 2017, \mn@doi [Astronomy & Astrophysics]
  {10.1051/0004-6361/201730533}, 603, A12

\bibitem[\protect\citeauthoryear{Bautista et~al.}{Bautista
  et~al.}{2020}]{eBOSS2020}
Bautista J.~E.,  et~al., 2020, \mn@doi [Mon. Not. Roy. Astron. Soc.]
  {10.1093/mnras/staa2800}, 500, 736

\bibitem[\protect\citeauthoryear{{Blanton} et~al.,}{{Blanton}
  et~al.}{2017}]{sdss4}
{Blanton} M.~R.,  et~al., 2017, \mn@doi [\aj] {10.3847/1538-3881/aa7567}, \href
  {https://ui.adsabs.harvard.edu/abs/2017AJ....154...28B} {154, 28}

\bibitem[\protect\citeauthoryear{{Blas}, {Lesgourgues}  \& {Tram}}{{Blas}
  et~al.}{2011}]{class2011}
{Blas} D.,  {Lesgourgues} J.,   {Tram} T.,  2011, \mn@doi [\jcap]
  {10.1088/1475-7516/2011/07/034}, \href
  {https://ui.adsabs.harvard.edu/abs/2011JCAP...07..034B} {2011, 034}

\bibitem[\protect\citeauthoryear{Bolton \& Becker}{Bolton \&
  Becker}{2009}]{bolton_resolving_2009}
Bolton J.~S.,  Becker G.~D.,  2009, \mn@doi [Monthly Notices of the Royal
  Astronomical Society: Letters] {10.1111/j.1745-3933.2009.00700.x}, 398, L26

\bibitem[\protect\citeauthoryear{{Bolton}, {Becker}, {Haehnelt}  \&
  {Viel}}{{Bolton} et~al.}{2014}]{Bolton2014}
{Bolton} J.~S.,  {Becker} G.~D.,  {Haehnelt} M.~G.,   {Viel} M.,  2014, \mn@doi
  [\mnras] {10.1093/mnras/stt2374}, \href
  {https://ui.adsabs.harvard.edu/abs/2014MNRAS.438.2499B} {438, 2499}

\bibitem[\protect\citeauthoryear{{Bolton}, {Puchwein}, {Sijacki}, {Haehnelt},
  {Kim}, {Meiksin}, {Regan}  \& {Viel}}{{Bolton} et~al.}{2017}]{Sherwood}
{Bolton} J.~S.,  {Puchwein} E.,  {Sijacki} D.,  {Haehnelt} M.~G.,  {Kim} T.-S.,
   {Meiksin} A.,  {Regan} J.~A.,   {Viel} M.,  2017, \mn@doi [\mnras]
  {10.1093/mnras/stw2397}, \href
  {https://ui.adsabs.harvard.edu/abs/2017MNRAS.464..897B} {464, 897}

\bibitem[\protect\citeauthoryear{Borde, Palanque-Delabrouille, Rossi, Viel,
  Bolton, Y{\`e}che, LeGoff  \& Rich}{Borde et~al.}{2014}]{borde_new_2014}
Borde A.,  Palanque-Delabrouille N.,  Rossi G.,  Viel M.,  Bolton J.,
  Y{\`e}che C.,  LeGoff J.-M.,   Rich J.,  2014, \mn@doi [Journal of Cosmology
  and Astroparticle Physics] {10.1088/1475-7516/2014/07/005}, 2014, 005

\bibitem[\protect\citeauthoryear{Bourboux et~al.,}{Bourboux
  et~al.}{2020}]{bourboux_completed_2020}
Bourboux H. d. M.~d.,  et~al., 2020, \mn@doi [ApJ] {10.3847/1538-4357/abb085},
  901, 153

\bibitem[\protect\citeauthoryear{{Chabanier} et~al.,}{{Chabanier}
  et~al.}{2019}]{Chabanier2019a}
{Chabanier} S.,  et~al., 2019, \mn@doi [\jcap] {10.1088/1475-7516/2019/07/017},
  \href {https://ui.adsabs.harvard.edu/abs/2019JCAP...07..017C} {2019, 017}

\bibitem[\protect\citeauthoryear{{Chabanier}, {Bournaud}, {Dubois},
  {Palanque-Delabrouille}, {Y{\`e}che}, {Armengaud}, {Peirani}  \&
  {Beckmann}}{{Chabanier} et~al.}{2020}]{Chabanier2020b}
{Chabanier} S.,  {Bournaud} F.,  {Dubois} Y.,  {Palanque-Delabrouille} N.,
  {Y{\`e}che} C.,  {Armengaud} E.,  {Peirani} S.,   {Beckmann} R.,  2020,
  \mn@doi [\mnras] {10.1093/mnras/staa1242}, \href
  {https://ui.adsabs.harvard.edu/abs/2020MNRAS.495.1825C} {495, 1825}

\bibitem[\protect\citeauthoryear{{Chabanier} et~al.,}{{Chabanier}
  et~al.}{2023}]{Chabanier2023}
{Chabanier} S.,  et~al., 2023, \mn@doi [\mnras] {10.1093/mnras/stac3294}, \href
  {https://ui.adsabs.harvard.edu/abs/2023MNRAS.518.3754C} {518, 3754}

\bibitem[\protect\citeauthoryear{{Chaussidon} et~al.,}{{Chaussidon}
  et~al.}{2022}]{Chaussidon2022}
{Chaussidon} E.,  et~al., 2022, arXiv e-prints, \href
  {https://ui.adsabs.harvard.edu/abs/2022arXiv220808511C} {p. arXiv:2208.08511}

\bibitem[\protect\citeauthoryear{Chen, Vlah  \& White}{Chen
  et~al.}{2021}]{chen_lyalpha_2021}
Chen S.-F.,  Vlah Z.,   White M.,  2021, arXiv:2103.13498 [astro-ph]

\bibitem[\protect\citeauthoryear{{Chisari} et~al.,}{{Chisari}
  et~al.}{2019}]{Chisari2019}
{Chisari} N.~E.,  et~al., 2019, \mn@doi [The Open Journal of Astrophysics]
  {10.21105/astro.1905.06082}, \href
  {https://ui.adsabs.harvard.edu/abs/2019OJAp....2E...4C} {2, 4}

\bibitem[\protect\citeauthoryear{Colella \& Woodward}{Colella \&
  Woodward}{1984}]{ppm}
Colella P.,  Woodward P.,  1984, \mn@doi [Journal of Computational Physics]
  {10.1016/0021-9991(84)90143-8}, 54, 174

\bibitem[\protect\citeauthoryear{{Cuceu}, {Font-Ribera}, {Nadathur}, {Joachimi}
   \& {Martini}}{{Cuceu} et~al.}{2022}]{Cuceu2022}
{Cuceu} A.,  {Font-Ribera} A.,  {Nadathur} S.,  {Joachimi} B.,   {Martini} P.,
  2022, arXiv e-prints, \href
  {https://ui.adsabs.harvard.edu/abs/2022arXiv220913942C} {p. arXiv:2209.13942}

\bibitem[\protect\citeauthoryear{DESI}{DESI}{2024}]{DESIlya2024}
DESI 2024, arXiv

\bibitem[\protect\citeauthoryear{{DESI Collaboration} et~al.,}{{DESI
  Collaboration} et~al.}{2016}]{DESI2016}
{DESI Collaboration} et~al., 2016, arXiv e-prints, \href
  {https://ui.adsabs.harvard.edu/abs/2016arXiv161100036D} {p. arXiv:1611.00036}

\bibitem[\protect\citeauthoryear{{Dawson} et~al.,}{{Dawson}
  et~al.}{2013}]{Dawson2013}
{Dawson} K.~S.,  et~al., 2013, \mn@doi [\aj] {10.1088/0004-6256/145/1/10},
  \href {https://ui.adsabs.harvard.edu/abs/2013AJ....145...10D} {145, 10}

\bibitem[\protect\citeauthoryear{{Dawson} et~al.,}{{Dawson}
  et~al.}{2016}]{Dawson2016}
{Dawson} K.~S.,  et~al., 2016, \mn@doi [\aj] {10.3847/0004-6256/151/2/44},
  \href {https://ui.adsabs.harvard.edu/abs/2016AJ....151...44D} {151, 44}

\bibitem[\protect\citeauthoryear{Dembinski, Ongmongkolkul  \& Deil}{Dembinski
  et~al.}{2022}]{dembinski_2022}
Dembinski H.,  Ongmongkolkul P.,   Deil C.,  2022, scikit-hep/iminuit: v2.15.0,
  \mn@doi{10.5281/zenodo.6916242}, \url
  {https://doi.org/10.5281/zenodo.6916242}

\bibitem[\protect\citeauthoryear{Doughty, Hennawi, Davies, Lukić  \&
  Oñorbe}{Doughty et~al.}{2023}]{doughty_convergence_2023}
Doughty C.~C.,  Hennawi J.~F.,  Davies F.~B.,  Lukić Z.,   Oñorbe J.,  2023,
  \mn@doi [Monthly Notices of the Royal Astronomical Society]
  {10.1093/mnras/stad2549}, 525, 3790

\bibitem[\protect\citeauthoryear{{Dvorkin}, {Blum}  \&
  {Kamionkowski}}{{Dvorkin} et~al.}{2014}]{Dvorkin2014}
{Dvorkin} C.,  {Blum} K.,   {Kamionkowski} M.,  2014, \mn@doi [\prd]
  {10.1103/PhysRevD.89.023519}, \href
  {https://ui.adsabs.harvard.edu/abs/2014PhRvD..89b3519D} {89, 023519}

\bibitem[\protect\citeauthoryear{{Eisenstein} \& {Hu}}{{Eisenstein} \&
  {Hu}}{1999}]{Einsenstein1999}
{Eisenstein} D.~J.,  {Hu} W.,  1999, \mn@doi [\apj] {10.1086/306640}, \href
  {https://ui.adsabs.harvard.edu/abs/1999ApJ...511....5E} {511, 5}

\bibitem[\protect\citeauthoryear{Eisenstein, Seo  \& White}{Eisenstein
  et~al.}{2007}]{Eisenstein2006}
Eisenstein D.~J.,  Seo H.-j.,   White M.~J.,  2007, \mn@doi [Astrophys. J.]
  {10.1086/518755}, 664, 660

\bibitem[\protect\citeauthoryear{{Eisenstein} et~al.,}{{Eisenstein}
  et~al.}{2011}]{sdss3}
{Eisenstein} D.~J.,  et~al., 2011, \mn@doi [\aj] {10.1088/0004-6256/142/3/72},
  \href {https://ui.adsabs.harvard.edu/abs/2011AJ....142...72E} {142, 72}

\bibitem[\protect\citeauthoryear{{Font-Ribera}, {McDonald}  \&
  {Slosar}}{{Font-Ribera} et~al.}{2018}]{FontRibera2018}
{Font-Ribera} A.,  {McDonald} P.,   {Slosar} A.,  2018, \mn@doi [\jcap]
  {10.1088/1475-7516/2018/01/003}, \href
  {https://ui.adsabs.harvard.edu/abs/2018JCAP...01..003F} {2018, 003}

\bibitem[\protect\citeauthoryear{{Friesen}, {Almgren}, {Luki{\'c}}, {Weber},
  {Morozov}, {Beckner}  \& {Day}}{{Friesen} et~al.}{2016}]{Briesen2016}
{Friesen} B.,  {Almgren} A.,  {Luki{\'c}} Z.,  {Weber} G.,  {Morozov} D.,
  {Beckner} V.,   {Day} M.,  2016, \mn@doi [Computational Astrophysics and
  Cosmology] {10.1186/s40668-016-0017-2}, \href
  {https://ui.adsabs.harvard.edu/abs/2016ComAC...3....4F} {3, 4}

\bibitem[\protect\citeauthoryear{{Garny}, {Konstandin}, {Sagunski}  \&
  {Tulin}}{{Garny} et~al.}{2018}]{Garny2018}
{Garny} M.,  {Konstandin} T.,  {Sagunski} L.,   {Tulin} S.,  2018, \mn@doi
  [\jcap] {10.1088/1475-7516/2018/09/011}, \href
  {https://ui.adsabs.harvard.edu/abs/2018JCAP...09..011G} {2018, 011}

\bibitem[\protect\citeauthoryear{Givans, Font-Ribera, Slosar, Seeyave,
  Pedersen, Rogers, Blas  \& Ir{\v s}i{\v c}}{Givans
  et~al.}{2022}]{givans_non-linearities_2022}
Givans J.~J.,  Font-Ribera A.,  Slosar A.,  Seeyave L.,  Pedersen C.,  Rogers
  K.~K.,  Blas D.,   Ir{\v s}i{\v c} V.,  2022, arXiv:2205.00962 [astro-ph]

\bibitem[\protect\citeauthoryear{Hadzhiyska et~al.}{Hadzhiyska
  et~al.}{2023}]{Hadzhiyska2023}
Hadzhiyska B.,  et~al., 2023, \mn@doi [Mon. Not. Roy. Astron. Soc.]
  {10.1093/mnras/stad1920}, 524, 1008

\bibitem[\protect\citeauthoryear{{Hahn}, {Rampf}  \& {Uhlemann}}{{Hahn}
  et~al.}{2021}]{Hahn2021}
{Hahn} O.,  {Rampf} C.,   {Uhlemann} C.,  2021, \mn@doi [\mnras]
  {10.1093/mnras/staa3773}, \href
  {https://ui.adsabs.harvard.edu/abs/2021MNRAS.503..426H} {503, 426}

\bibitem[\protect\citeauthoryear{{Hern{\'a}ndez-Aguayo}
  et~al.,}{{Hern{\'a}ndez-Aguayo} et~al.}{2022}]{MilleniumTNG}
{Hern{\'a}ndez-Aguayo} C.,  et~al., 2022, arXiv e-prints, \href
  {https://ui.adsabs.harvard.edu/abs/2022arXiv221010059H} {p. arXiv:2210.10059}

\bibitem[\protect\citeauthoryear{Ir{\v s}i{\v c}, Viel, Haehnelt, Bolton  \&
  Becker}{Ir{\v s}i{\v c} et~al.}{2017}]{irsic_first_2017}
Ir{\v s}i{\v c} V.,  Viel M.,  Haehnelt M.~G.,  Bolton J.~S.,   Becker G.~D.,
  2017, \mn@doi [Physical Review Letters] {10.1103/PhysRevLett.119.031302},
  119, 031302

\bibitem[\protect\citeauthoryear{Iršič et~al.,}{Iršič
  et~al.}{2024}]{irsic_unveiling_2024}
Iršič V.,  et~al., 2024, \mn@doi [Phys. Rev. D]
  {10.1103/PhysRevD.109.043511}, 109, 043511

\bibitem[\protect\citeauthoryear{Ivanov}{Ivanov}{2024}]{Ivanov2023}
Ivanov M.~M.,  2024, \mn@doi [Phys. Rev. D] {10.1103/PhysRevD.109.023507}, 109,
  023507

\bibitem[\protect\citeauthoryear{{Kaiser}}{{Kaiser}}{1987}]{Kaiser1987}
{Kaiser} N.,  1987, \mn@doi [\mnras] {10.1093/mnras/227.1.1}, \href
  {https://ui.adsabs.harvard.edu/abs/1987MNRAS.227....1K} {227, 1}

\bibitem[\protect\citeauthoryear{{Kara{\c{c}}ayl{\i}}
  et~al.,}{{Kara{\c{c}}ayl{\i}} et~al.}{2022}]{Naim2022}
{Kara{\c{c}}ayl{\i}} N.~G.,  et~al., 2022, \mn@doi [\mnras]
  {10.1093/mnras/stab3201}, \href
  {https://ui.adsabs.harvard.edu/abs/2022MNRAS.509.2842K} {509, 2842}

\bibitem[\protect\citeauthoryear{Kara\c{c}ayl\i{} et~al.}{Kara\c{c}ayl\i{}
  et~al.}{2024}]{Karacayli2023}
Kara\c{c}ayl\i{} N.~G.,  et~al., 2024, \mn@doi [Mon. Not. Roy. Astron. Soc.]
  {10.1093/mnras/stae171}, 528, 3941

\bibitem[\protect\citeauthoryear{Kirkby et~al.,}{Kirkby
  et~al.}{2013}]{kirkby_fitting_2013}
Kirkby D.,  et~al., 2013, \mn@doi [J. Cosmol. Astropart. Phys.]
  {10.1088/1475-7516/2013/03/024}, 2013, 024

\bibitem[\protect\citeauthoryear{{Kulkarni}, {Hennawi}, {O{\~n}orbe}, {Rorai}
  \& {Springel}}{{Kulkarni} et~al.}{2015}]{Kulkarni2015}
{Kulkarni} G.,  {Hennawi} J.~F.,  {O{\~n}orbe} J.,  {Rorai} A.,   {Springel}
  V.,  2015, \mn@doi [\apj] {10.1088/0004-637X/812/1/30}, \href
  {https://ui.adsabs.harvard.edu/abs/2015ApJ...812...30K} {812, 30}

\bibitem[\protect\citeauthoryear{{Luki{\'c}}, {Stark}, {Nugent}, {White},
  {Meiksin}  \& {Almgren}}{{Luki{\'c}} et~al.}{2015}]{Lukic2015}
{Luki{\'c}} Z.,  {Stark} C.~W.,  {Nugent} P.,  {White} M.,  {Meiksin} A.~A.,
  {Almgren} A.,  2015, \mn@doi [\mnras] {10.1093/mnras/stu2377}, \href
  {https://ui.adsabs.harvard.edu/abs/2015MNRAS.446.3697L} {446, 3697}

\bibitem[\protect\citeauthoryear{McDonald}{McDonald}{2003}]{mcdonald_toward_2003}
McDonald P.,  2003, \mn@doi [The Astrophysical Journal] {10.1086/345945}, 585,
  34

\bibitem[\protect\citeauthoryear{{O{\~n}orbe}, {Hennawi}  \&
  {Luki{\'c}}}{{O{\~n}orbe} et~al.}{2017}]{Onorbe2017}
{O{\~n}orbe} J.,  {Hennawi} J.~F.,   {Luki{\'c}} Z.,  2017, \mn@doi [\apj]
  {10.3847/1538-4357/aa6031}, \href
  {https://ui.adsabs.harvard.edu/abs/2017ApJ...837..106O} {837, 106}

\bibitem[\protect\citeauthoryear{Palanque-Delabrouille
  et~al.,}{Palanque-Delabrouille et~al.}{2013}]{PalanqueDelabrouille2013}
Palanque-Delabrouille N.,  et~al., 2013, \mn@doi [Astronomy & Astrophysics]
  {10.1051/0004-6361/201322130}, 559, A85

\bibitem[\protect\citeauthoryear{Palanque-Delabrouille
  et~al.,}{Palanque-Delabrouille et~al.}{2015a}]{PalanqueDelabrouille2015a}
Palanque-Delabrouille N.,  et~al., 2015a, \mn@doi [Journal of Cosmology and
  Astroparticle Physics] {10.1088/1475-7516/2015/02/045}, 2015, 045–045

\bibitem[\protect\citeauthoryear{Palanque-Delabrouille
  et~al.,}{Palanque-Delabrouille et~al.}{2015b}]{PalanqueDelabrouille2015b}
Palanque-Delabrouille N.,  et~al., 2015b, \mn@doi [Journal of Cosmology and
  Astroparticle Physics] {10.1088/1475-7516/2015/11/011}, 2015, 011–011

\bibitem[\protect\citeauthoryear{{Palanque-Delabrouille}
  et~al.,}{{Palanque-Delabrouille} et~al.}{2016}]{PalanqueDelabrouille2016}
{Palanque-Delabrouille} N.,  et~al., 2016, \mn@doi [\aap]
  {10.1051/0004-6361/201527392}, \href
  {https://ui.adsabs.harvard.edu/abs/2016A&A...587A..41P} {587, A41}

\bibitem[\protect\citeauthoryear{{Palanque-Delabrouille}, {Y{\`e}che},
  {Sch{\"o}neberg}, {Lesgourgues}, {Walther}, {Chabanier}  \&
  {Armengaud}}{{Palanque-Delabrouille} et~al.}{2019}]{PDB2019}
{Palanque-Delabrouille} N.,  {Y{\`e}che} C.,  {Sch{\"o}neberg} N.,
  {Lesgourgues} J.,  {Walther} M.,  {Chabanier} S.,   {Armengaud} E.,  2019,
  arXiv e-prints, \href {https://ui.adsabs.harvard.edu/abs/2019arXiv191109073P}
  {p. arXiv:1911.09073}

\bibitem[\protect\citeauthoryear{{Pedersen}, {Font-Ribera}, {Rogers},
  {McDonald}, {Peiris}, {Pontzen}  \& {Slosar}}{{Pedersen}
  et~al.}{2021}]{Pedersen2021}
{Pedersen} C.,  {Font-Ribera} A.,  {Rogers} K.~K.,  {McDonald} P.,  {Peiris}
  H.~V.,  {Pontzen} A.,   {Slosar} A.,  2021, \mn@doi [\jcap]
  {10.1088/1475-7516/2021/05/033}, \href
  {https://ui.adsabs.harvard.edu/abs/2021JCAP...05..033P} {2021, 033}

\bibitem[\protect\citeauthoryear{{Peeples}, {Weinberg}, {Dav{\'e}}, {Fardal}
  \& {Katz}}{{Peeples} et~al.}{2010}]{Peeples2010}
{Peeples} M.~S.,  {Weinberg} D.~H.,  {Dav{\'e}} R.,  {Fardal} M.~A.,   {Katz}
  N.,  2010, \mn@doi [\mnras] {10.1111/j.1365-2966.2010.16383.x}, \href
  {https://ui.adsabs.harvard.edu/abs/2010MNRAS.404.1281P} {404, 1281}

\bibitem[\protect\citeauthoryear{Pieri et~al.}{Pieri et~al.}{2016}]{WEAVE2016}
Pieri M.~M.,  et~al., 2016.  (\mn@eprint {arXiv} {1611.09388})

\bibitem[\protect\citeauthoryear{{Pillepich} et~al.,}{{Pillepich}
  et~al.}{2019}]{Pillepich2019}
{Pillepich} A.,  et~al., 2019, \mn@doi [\mnras] {10.1093/mnras/stz2338}, \href
  {https://ui.adsabs.harvard.edu/abs/2019MNRAS.490.3196P} {490, 3196}

\bibitem[\protect\citeauthoryear{{Planck Collaboration} et~al.,}{{Planck
  Collaboration} et~al.}{2016}]{Planck2015}
{Planck Collaboration} et~al., 2016, \mn@doi [\aap]
  {10.1051/0004-6361/201525830}, \href
  {https://ui.adsabs.harvard.edu/abs/2016A&A...594A..13P} {594, A13}

\bibitem[\protect\citeauthoryear{Ravoux et~al.}{Ravoux
  et~al.}{2023}]{Ravoux2023}
Ravoux C.,  et~al., 2023, \mn@doi [Mon. Not. Roy. Astron. Soc.]
  {10.1093/mnras/stad3008}, 526, 5118

\bibitem[\protect\citeauthoryear{Rogers \& Peiris}{Rogers \&
  Peiris}{2021}]{rogers_strong_2021}
Rogers K.~K.,  Peiris H.~V.,  2021, \mn@doi [Phys. Rev. Lett.]
  {10.1103/PhysRevLett.126.071302}, 126, 071302

\bibitem[\protect\citeauthoryear{Schaye et~al.}{Schaye
  et~al.}{2023}]{Schaye2023}
Schaye J.,  et~al., 2023, \mn@doi [Mon. Not. Roy. Astron. Soc.]
  {10.1093/mnras/stad2419}, 526, 4978

\bibitem[\protect\citeauthoryear{{Seljak} et~al.,}{{Seljak}
  et~al.}{2005}]{Seljak2005}
{Seljak} U.,  et~al., 2005, \mn@doi [\prd] {10.1103/PhysRevD.71.103515}, \href
  {https://ui.adsabs.harvard.edu/abs/2005PhRvD..71j3515S} {71, 103515}

\bibitem[\protect\citeauthoryear{Seo \& Eisenstein}{Seo \&
  Eisenstein}{2007}]{Seo2007}
Seo H.-J.,  Eisenstein D.~J.,  2007, \mn@doi [Astrophys. J.] {10.1086/519549},
  665, 14

\bibitem[\protect\citeauthoryear{Sexton, Luki\'c, Almgren, Daley, Friesen,
  Myers  \& Zhang}{Sexton et~al.}{2021}]{Sexton2021}
Sexton J.,  Luki\'c Z.,  Almgren A.,  Daley C.,  Friesen B.,  Myers A.,   Zhang
  W.,  2021, \mn@doi [Journal of Open Source Software] {10.21105/joss.03068},
  6, 3068

\bibitem[\protect\citeauthoryear{Slosar et~al.,}{Slosar
  et~al.}{2013}]{Slosar2013}
Slosar A.,  et~al., 2013, \mn@doi [Journal of Cosmology and Astroparticle
  Physics] {10.1088/1475-7516/2013/04/026}, 2013, 026–026

\bibitem[\protect\citeauthoryear{Viel, Lesgourgues, Haehnelt, Matarrese  \&
  Riotto}{Viel et~al.}{2005}]{viel_constraining_2005}
Viel M.,  Lesgourgues J.,  Haehnelt M.~G.,  Matarrese S.,   Riotto A.,  2005,
  \mn@doi [Physical Review D] {10.1103/PhysRevD.71.063534}, 71, 063534

\bibitem[\protect\citeauthoryear{Viel, Becker, Bolton  \& Haehnelt}{Viel
  et~al.}{2013}]{viel_warm_2013}
Viel M.,  Becker G.~D.,  Bolton J.~S.,   Haehnelt M.~G.,  2013, \mn@doi [Phys.
  Rev. D] {10.1103/PhysRevD.88.043502}, 88, 043502

\bibitem[\protect\citeauthoryear{Villasenor, Robertson, Madau  \&
  Schneider}{Villasenor et~al.}{2023}]{villasenor_new_2023}
Villasenor B.,  Robertson B.,  Madau P.,   Schneider E.,  2023, \mn@doi [Phys.
  Rev. D] {10.1103/PhysRevD.108.023502}, 108, 023502

\bibitem[\protect\citeauthoryear{Wallisch}{Wallisch}{2018}]{Wallisch2018}
Wallisch B.,  2018, PhD thesis, Cambridge U. (\mn@eprint {arXiv} {1810.02800}),
  \mn@doi{10.17863/CAM.30368}

\bibitem[\protect\citeauthoryear{{Walther}, {Hennawi}, {Hiss}, {O{\~n}orbe},
  {Lee}, {Rorai}  \& {O'Meara}}{{Walther} et~al.}{2018}]{Walther2018}
{Walther} M.,  {Hennawi} J.~F.,  {Hiss} H.,  {O{\~n}orbe} J.,  {Lee} K.-G.,
  {Rorai} A.,   {O'Meara} J.,  2018, \mn@doi [\apj] {10.3847/1538-4357/aa9c81},
  \href {https://ui.adsabs.harvard.edu/abs/2018ApJ...852...22W} {852, 22}

\bibitem[\protect\citeauthoryear{{Walther}, {O{\~n}orbe}, {Hennawi}  \&
  {Luki{\'c}}}{{Walther} et~al.}{2019}]{Walther2019}
{Walther} M.,  {O{\~n}orbe} J.,  {Hennawi} J.~F.,   {Luki{\'c}} Z.,  2019,
  \mn@doi [\apj] {10.3847/1538-4357/aafad1}, \href
  {https://ui.adsabs.harvard.edu/abs/2019ApJ...872...13W} {872, 13}

\bibitem[\protect\citeauthoryear{{Walther}, {Armengaud}, {Ravoux},
  {Palanque-Delabrouille}, {Y{\`e}che}  \& {Luki{\'c}}}{{Walther}
  et~al.}{2021}]{Walther2021}
{Walther} M.,  {Armengaud} E.,  {Ravoux} C.,  {Palanque-Delabrouille} N.,
  {Y{\`e}che} C.,   {Luki{\'c}} Z.,  2021, \mn@doi [\jcap]
  {10.1088/1475-7516/2021/04/059}, \href
  {https://ui.adsabs.harvard.edu/abs/2021JCAP...04..059W} {2021, 059}

\bibitem[\protect\citeauthoryear{{Xu}, {Dvorkin}  \& {Chael}}{{Xu}
  et~al.}{2018}]{Xu2018}
{Xu} W.~L.,  {Dvorkin} C.,   {Chael} A.,  2018, \mn@doi [\prd]
  {10.1103/PhysRevD.97.103530}, \href
  {https://ui.adsabs.harvard.edu/abs/2018PhRvD..97j3530X} {97, 103530}

\bibitem[\protect\citeauthoryear{Y\`eche, Palanque-Delabrouille, Baur  \&
  Bourboux}{Y\`eche et~al.}{2017}]{Yeche2017}
Y\`eche C.,  Palanque-Delabrouille N.,  Baur J.,   Bourboux H. d. M.~d.,  2017,
  \mn@doi [Journal of Cosmology and Astroparticle Physics]
  {10.1088/1475-7516/2017/06/047}, 2017, 047–047

\bibitem[\protect\citeauthoryear{{Zel'dovich}}{{Zel'dovich}}{1970}]{zeldovich1970}
{Zel'dovich} Y.~B.,  1970, \aap, \href
  {https://ui.adsabs.harvard.edu/abs/1970A&A.....5...84Z} {5, 84}

\bibitem[\protect\citeauthoryear{Zhang et~al.,}{Zhang et~al.}{2019}]{AMReX}
Zhang W.,  et~al., 2019, \mn@doi [Journal of Open Source Software]
  {10.21105/joss.01370}, 4, 1370

\bibitem[\protect\citeauthoryear{de Belsunce, Philcox, Irsic, McDonald, Guy  \&
  Palanque-Delabrouille}{de~Belsunce et~al.}{2024}]{Belsunce2024}
de Belsunce R.,  Philcox O. H.~E.,  Irsic V.,  McDonald P.,  Guy J.,
  Palanque-Delabrouille N.,  2024, arXiv

\bibitem[\protect\citeauthoryear{de Sainte~Agathe et~al.,}{de~Sainte~Agathe
  et~al.}{2019}]{deSainteAgathe2019}
de Sainte~Agathe V.,  et~al., 2019, \mn@doi [Astronomy & Astrophysics]
  {10.1051/0004-6361/201935638}, 629, A85

\bibitem[\protect\citeauthoryear{{du Mas des Bourboux} et~al.,}{{du Mas des
  Bourboux} et~al.}{2020}]{DuMas2020}
{du Mas des Bourboux} H.,  et~al., 2020, \mn@doi [\apj]
  {10.3847/1538-4357/abb085}, \href
  {https://ui.adsabs.harvard.edu/abs/2020ApJ...901..153D} {901, 153}

\makeatother
\end{thebibliography}




\appendix

\section{Detail of fitted linear and non-linear parameters}
\label{appendix:param_nosplice}

\begin{table*}
	\centering
    \caption{Table of all the fitted parameters for non-spliced \pkt~from the \textit{L160R100}, \textit{L160R50}, and \textit{L320R100} simulations.\com{ The $D_1$ parameterization (Equation~\ref{eq:model_p3d_nl_1}) is used here.}}
    \label{tab:results_no_splice_1}
    \begin{tabular}{lccccc}
    \hline\hline
    \multicolumn{5}{c}{Grid $ACCL2\_L160R100$} \\
    \hline\hline
    $z$ & 4.0 & 3.6 & 3.0 & 2.6 & 2.0  \\
    \hline
    $b_{\alpha}$ & -0.53 $\pm$ 0.00781 & -0.394 $\pm$ 0.00599 & -0.233 $\pm$ 0.00384 & -0.151 $\pm$ 0.00339 & -0.0703 $\pm$ 0.00162  \\
    $\beta_{\alpha}$ & 0.346 $\pm $ 0.0459 & 0.478 $\pm $ 0.0495 & 0.781 $\pm $ 0.0584 & 1.0 $\pm $ 0.0679 & 1.28 $\pm $ 0.0793  \\
    $q_1$ & 2.52 $\pm$ 0.102 & 2.18 $\pm$ 0.0893 & 1.94 $\pm$ 0.0742 & 1.9 $\pm$ 0.14 & 1.6 $\pm$ 0.102  \\
    $q_2$ & -1.0 $\pm$ 0.0579 & -1.0 $\pm$ 0.0482 & -1.0 $\pm$ 0.0553 & -0.879 $\pm$ 0.112 & -0.416 $\pm$ 0.0586  \\
    $k_v$ & 16.7 $\pm$ 7.29 & 8.45 $\pm$ 2.01 & 3.82 $\pm$ 0.505 & 3.37 $\pm$ 0.345 & 3.14 $\pm$ 0.299  \\
    $a_v$ & 0.252 $\pm$ 0.0993 & 0.314 $\pm$ 0.0975 & 0.51 $\pm$ 0.0989 & 0.586 $\pm$ 0.0949 & 0.471 $\pm$ 0.0634  \\
    $b_v$ & 1.91 $\pm$ 0.222 & 1.87 $\pm$ 0.213 & 1.82 $\pm$ 0.18 & 1.74 $\pm$ 0.147 & 1.51 $\pm$ 0.0899  \\
    $k_p$ & 7.11 $\pm$ 0.146 & 7.41 $\pm$ 0.165 & 8.35 $\pm$ 0.246 & 9.13 $\pm$ 0.447 & 9.03 $\pm$ 0.465  \\
    $\chi^{2}$ & 112.0 & 121.0 & 146.0 & 168.0 & 182.0  \\
    Reduced $\chi^{2}$ & 1.49 & 1.61 & 1.95 & 2.24 & 2.42  \\
    \hline\hline
    \multicolumn{5}{c}{Grid $ACCL2\_L160R50$} \\
    \hline\hline
    $z$ & 4.0 & 3.6 & 3.0 & 2.6 & 2.0  \\
    \hline
    $b_{\alpha}$ & -0.459 $\pm$ 0.00717 & -0.347 $\pm$ 0.0057 & -0.217 $\pm$ 0.00468 & -0.15 $\pm$ 0.0033 & -0.0758 $\pm$ 0.00178  \\
    $\beta_{\alpha}$ & 0.513 $\pm $ 0.0514 & 0.702 $\pm $ 0.0565 & 1.06 $\pm $ 0.0659 & 1.27 $\pm $ 0.0722 & 1.51 $\pm $ 0.0843  \\
    $q_1$ & 2.5 $\pm$ 0.112 & 2.14 $\pm$ 0.0999 & 1.62 $\pm$ 0.161 & 1.33 $\pm$ 0.132 & 1.11 $\pm$ 0.104  \\
    $q_2$ & -1.0 $\pm$ 0.0965 & -1.0 $\pm$ 0.109 & -0.72 $\pm$ 0.163 & -0.426 $\pm$ 0.11 & -0.136 $\pm$ 0.0624  \\
    $k_v$ & 6.29 $\pm$ 1.1 & 3.59 $\pm$ 0.551 & 2.23 $\pm$ 0.292 & 1.81 $\pm$ 0.247 & 1.37 $\pm$ 0.217  \\
    $a_v$ & 0.339 $\pm$ 0.0932 & 0.445 $\pm$ 0.0938 & 0.617 $\pm$ 0.103 & 0.592 $\pm$ 0.0891 & 0.432 $\pm$ 0.058  \\
    $b_v$ & 1.63 $\pm$ 0.162 & 1.64 $\pm$ 0.148 & 1.68 $\pm$ 0.123 & 1.63 $\pm$ 0.1 & 1.45 $\pm$ 0.0646  \\
    $k_p$ & 11.7 $\pm$ 0.714 & 13.4 $\pm$ 1.1 & 17.8 $\pm$ 3.57 & 19.5 $\pm$ 4.83 & 17.3 $\pm$ 3.68  \\
    $\chi^{2}$ & 100.0 & 109.0 & 130.0 & 141.0 & 159.0  \\
    Reduced $\chi^{2}$ & 1.33 & 1.46 & 1.73 & 1.88 & 2.12  \\
    \hline\hline
    \multicolumn{5}{c}{Grid $ACCL2\_L320R100$} \\
    \hline\hline
    $z$ & 4.0 & 3.6 & 3.0 & 2.6 & 2.0  \\
    \hline
    $b_{\alpha}$ & -0.544 $\pm$ 0.00595 & -0.405 $\pm$ 0.00453 & -0.244 $\pm$ 0.00355 & -0.163 $\pm$ 0.00248 & -0.0758 $\pm$ 0.00121  \\
    $\beta_{\alpha}$ & 0.363 $\pm $ 0.0362 & 0.502 $\pm $ 0.0391 & 0.805 $\pm $ 0.0443 & 1.03 $\pm $ 0.0516 & 1.36 $\pm $ 0.061  \\
    $q_1$ & 2.34 $\pm$ 0.0683 & 2.03 $\pm$ 0.0582 & 1.63 $\pm$ 0.115 & 1.39 $\pm$ 0.0979 & 1.23 $\pm$ 0.0755  \\
    $q_2$ & -1.0 $\pm$ 0.0429 & -1.0 $\pm$ 0.0462 & -0.78 $\pm$ 0.117 & -0.503 $\pm$ 0.0776 & -0.217 $\pm$ 0.0425  \\
    $k_v$ & 12.3 $\pm$ 3.53 & 5.56 $\pm$ 1.21 & 2.56 $\pm$ 0.448 & 2.22 $\pm$ 0.348 & 1.87 $\pm$ 0.306  \\
    $a_v$ & 0.208 $\pm$ 0.0731 & 0.253 $\pm$ 0.0704 & 0.418 $\pm$ 0.0733 & 0.432 $\pm$ 0.0701 & 0.31 $\pm$ 0.0431  \\
    $b_v$ & 2.0 $\pm$ 0.13 & 2.0 $\pm$ 0.161 & 2.0 $\pm$ 0.276 & 1.89 $\pm$ 0.134 & 1.51 $\pm$ 0.075  \\
    $k_p$ & 7.29 $\pm$ 0.0904 & 7.53 $\pm$ 0.0988 & 8.07 $\pm$ 0.173 & 8.26 $\pm$ 0.191 & 8.12 $\pm$ 0.195  \\
    $\chi^{2}$ & 80.7 & 79.1 & 84.7 & 85.7 & 79.8  \\
    Reduced $\chi^{2}$ & 0.917 & 0.899 & 0.962 & 0.974 & 0.906  \\
    \hline\hline
    \end{tabular}
\end{table*}

\begin{table*}
	\centering
    \caption{Table of all the fitted parameters for non-spliced \pkt~from the \textit{L160R25}, \textit{L320R50}, and \textit{L640R100} simulations\com{. The $D_1$ parameterization (Equation~\ref{eq:model_p3d_nl_1}) is used here.}}
    \label{tab:results_no_splice_2}
    \begin{tabular}{lccccc}
    \hline\hline
    \multicolumn{5}{c}{Grid $ACCL2\_L160R25$} \\
    \hline\hline
    $z$ & 4.0 & 3.6 & 3.0 & 2.6 & 2.0  \\
    \hline
    $b_{\alpha}$ & -0.404 $\pm$ 0.00676 & -0.315 $\pm$ 0.00701 & -0.209 $\pm$ 0.00453 & -0.15 $\pm$ 0.00334 & -0.0805 $\pm$ 0.002  \\
    $\beta_{\alpha}$ & 0.874 $\pm $ 0.0717 & 1.07 $\pm $ 0.0752 & 1.42 $\pm $ 0.0774 & 1.61 $\pm $ 0.0819 & 1.75 $\pm $ 0.0918  \\
    $q_1$ & 2.55 $\pm$ 0.121 & 1.92 $\pm$ 0.239 & 1.08 $\pm$ 0.171 & 0.779 $\pm$ 0.147 & 0.562 $\pm$ 0.125  \\
    $q_2$ & -1.0 $\pm$ 0.349 & -0.684 $\pm$ 0.33 & -0.156 $\pm$ 0.192 & 0.0235 $\pm$ 0.137 & 0.16 $\pm$ 0.0835  \\
    $k_v$ & 2.67 $\pm$ 0.856 & 1.78 $\pm$ 0.516 & 1.01 $\pm$ 0.324 & 0.61 $\pm$ 0.266 & 0.152 $\pm$ 0.15  \\
    $a_v$ & 0.25 $\pm$ 0.0781 & 0.353 $\pm$ 0.093 & 0.445 $\pm$ 0.1 & 0.394 $\pm$ 0.0916 & 0.225 $\pm$ 0.068  \\
    $b_v$ & 1.65 $\pm$ 0.131 & 1.69 $\pm$ 0.126 & 1.74 $\pm$ 0.109 & 1.7 $\pm$ 0.0907 & 1.54 $\pm$ 0.0613  \\
    $k_p$ & 16.0 $\pm$ 1.9 & 18.4 $\pm$ 4.38 & 21.0 $\pm$ 6.89 & 21.3 $\pm$ 7.61 & 15.8 $\pm$ 3.63  \\
    $\chi^{2}$ & 33.5 & 35.5 & 38.1 & 39.9 & 44.7  \\
    Reduced $\chi^{2}$ & 0.447 & 0.473 & 0.508 & 0.532 & 0.596  \\
    \hline\hline
    \multicolumn{5}{c}{Grid $ACCL2\_L320R50$} \\
    \hline\hline
    $z$ & 4.0 & 3.6 & 3.0 & 2.6 & 2.0  \\
    \hline
    $b_{\alpha}$ & -0.459 $\pm$ 0.00524 & -0.351 $\pm$ 0.00516 & -0.225 $\pm$ 0.0033 & -0.157 $\pm$ 0.00235 & -0.0797 $\pm$ 0.00128  \\
    $\beta_{\alpha}$ & 0.596 $\pm $ 0.0427 & 0.782 $\pm $ 0.0464 & 1.13 $\pm $ 0.0503 & 1.36 $\pm $ 0.0546 & 1.61 $\pm $ 0.0621  \\
    $q_1$ & 2.53 $\pm$ 0.0844 & 2.0 $\pm$ 0.166 & 1.22 $\pm$ 0.119 & 0.97 $\pm$ 0.1 & 0.814 $\pm$ 0.0797  \\
    $q_2$ & -1.0 $\pm$ 0.152 & -0.745 $\pm$ 0.221 & -0.228 $\pm$ 0.126 & -0.0674 $\pm$ 0.0873 & 0.0779 $\pm$ 0.0492  \\
    $k_v$ & 4.26 $\pm$ 0.922 & 2.42 $\pm$ 0.524 & 1.29 $\pm$ 0.297 & 0.906 $\pm$ 0.237 & 0.489 $\pm$ 0.174  \\
    $a_v$ & 0.238 $\pm$ 0.061 & 0.316 $\pm$ 0.0661 & 0.39 $\pm$ 0.0661 & 0.359 $\pm$ 0.0575 & 0.241 $\pm$ 0.0382  \\
    $b_v$ & 1.71 $\pm$ 0.128 & 1.71 $\pm$ 0.118 & 1.71 $\pm$ 0.0978 & 1.63 $\pm$ 0.0791 & 1.41 $\pm$ 0.0507  \\
    $k_p$ & 11.4 $\pm$ 0.385 & 12.0 $\pm$ 0.648 & 12.5 $\pm$ 0.754 & 12.6 $\pm$ 0.807 & 11.5 $\pm$ 0.662  \\
    $\chi^{2}$ & 39.4 & 39.6 & 40.3 & 39.7 & 44.9  \\
    Reduced $\chi^{2}$ & 0.448 & 0.45 & 0.459 & 0.451 & 0.51  \\
    \hline\hline
    \multicolumn{5}{c}{Grid $ACCL2\_L640R100$} \\
    \hline\hline
    $z$ & 4.0 & 3.6 & 3.0 & 2.6 & 2.0  \\
    \hline
    $b_{\alpha}$ & -0.543 $\pm$ 0.00439 & -0.406 $\pm$ 0.00336 & -0.249 $\pm$ 0.00256 & -0.165 $\pm$ 0.00171 & -0.0769 $\pm$ 0.000826  \\
    $\beta_{\alpha}$ & 0.433 $\pm $ 0.0277 & 0.57 $\pm $ 0.03 & 0.862 $\pm $ 0.0354 & 1.09 $\pm $ 0.0383 & 1.42 $\pm $ 0.0442  \\
    $q_1$ & 2.37 $\pm$ 0.0563 & 2.04 $\pm$ 0.0481 & 1.47 $\pm$ 0.0995 & 1.26 $\pm$ 0.0799 & 1.15 $\pm$ 0.0604  \\
    $q_2$ & -1.0 $\pm$ 0.0428 & -1.0 $\pm$ 0.063 & -0.607 $\pm$ 0.104 & -0.388 $\pm$ 0.0685 & -0.167 $\pm$ 0.0369  \\
    $k_v$ & 9.74 $\pm$ 4.11 & 2.95 $\pm$ 1.04 & 1.53 $\pm$ 0.39 & 1.44 $\pm$ 0.302 & 1.27 $\pm$ 0.251  \\
    $a_v$ & 0.107 $\pm$ 0.0955 & 0.155 $\pm$ 0.0507 & 0.291 $\pm$ 0.0596 & 0.316 $\pm$ 0.0532 & 0.241 $\pm$ 0.0339  \\
    $b_v$ & 2.0 $\pm$ 0.183 & 2.0 $\pm$ 1.21 & 1.97 $\pm$ 0.218 & 1.84 $\pm$ 0.123 & 1.51 $\pm$ 0.0722  \\
    $k_p$ & 7.19 $\pm$ 0.0805 & 7.43 $\pm$ 0.0886 & 7.8 $\pm$ 0.15 & 7.96 $\pm$ 0.162 & 7.85 $\pm$ 0.165  \\
    $\chi^{2}$ & 68.2 & 62.2 & 64.7 & 64.3 & 67.2  \\
    Reduced $\chi^{2}$ & 0.609 & 0.555 & 0.578 & 0.574 & 0.6  \\
    \hline\hline
    \end{tabular}
\end{table*}

\begin{table*}
	\centering
    \caption{Table of all the fitted parameters for different fits of the non-spliced \textit{L160R25} simulation at all redshifts using the non-linear parametrization D1, D1($q_2=0$) and D0.}
    \label{tab:results_no_splice_fit_variation}
    \begin{tabular}{lccccc}
    \hline\hline
    \multicolumn{5}{c}{Fit D1} \\
    \hline\hline
    $z$ & 4.0 & 3.6 & 3.0 & 2.6 & 2.0  \\
    \hline
    $b_{\alpha}$ & -0.404 $\pm$ 0.00676 & -0.315 $\pm$ 0.00701 & -0.209 $\pm$ 0.00453 & -0.15 $\pm$ 0.00334 & -0.0805 $\pm$ 0.002  \\
    $\beta_{\alpha}$ & 0.874 $\pm $ 0.0717 & 1.07 $\pm $ 0.0752 & 1.42 $\pm $ 0.0774 & 1.61 $\pm $ 0.0819 & 1.75 $\pm $ 0.0918  \\
    $q_1$ & 2.55 $\pm$ 0.121 & 1.92 $\pm$ 0.239 & 1.08 $\pm$ 0.171 & 0.779 $\pm$ 0.147 & 0.562 $\pm$ 0.125  \\
    $q_2$ & -1.0 $\pm$ 0.349 & -0.684 $\pm$ 0.33 & -0.156 $\pm$ 0.192 & 0.0235 $\pm$ 0.137 & 0.16 $\pm$ 0.0835  \\
    $k_v$ & 2.67 $\pm$ 0.856 & 1.78 $\pm$ 0.516 & 1.01 $\pm$ 0.324 & 0.61 $\pm$ 0.266 & 0.152 $\pm$ 0.15  \\
    $a_v$ & 0.25 $\pm$ 0.0781 & 0.353 $\pm$ 0.093 & 0.445 $\pm$ 0.1 & 0.394 $\pm$ 0.0916 & 0.225 $\pm$ 0.068  \\
    $b_v$ & 1.65 $\pm$ 0.131 & 1.69 $\pm$ 0.126 & 1.74 $\pm$ 0.109 & 1.7 $\pm$ 0.0907 & 1.54 $\pm$ 0.0613  \\
    $k_p$ & 16.0 $\pm$ 1.9 & 18.4 $\pm$ 4.38 & 21.0 $\pm$ 6.89 & 21.3 $\pm$ 7.61 & 15.8 $\pm$ 3.63  \\
    $\chi^{2}$ & 33.5 & 35.5 & 38.1 & 39.9 & 44.7  \\
    Reduced $\chi^{2}$ & 0.447 & 0.473 & 0.508 & 0.532 & 0.596  \\
    \hline\hline
    \multicolumn{5}{c}{Fit D1 ($q_2=0$)} \\
    \hline\hline
    $z$ & 4.0 & 3.6 & 3.0 & 2.6 & 2.0  \\
    \hline
    $b_{\alpha}$ & -0.416 $\pm$ 0.00672 & -0.323 $\pm$ 0.00545 & -0.211 $\pm$ 0.00378 & -0.149 $\pm$ 0.00276 & -0.0779 $\pm$ 0.0015  \\
    $\beta_{\alpha}$ & 0.854 $\pm $ 0.0669 & 1.06 $\pm $ 0.0705 & 1.42 $\pm $ 0.0761 & 1.61 $\pm $ 0.0821 & 1.77 $\pm $ 0.0951  \\
    $q_1$ & 2.02 $\pm$ 0.116 & 1.5 $\pm$ 0.101 & 0.957 $\pm$ 0.0786 & 0.801 $\pm$ 0.0657 & 0.786 $\pm$ 0.0494  \\
    $q_2$ & 0.0 $\pm$ 0.1 & 0.0 $\pm$ 0.1 & 0.0 $\pm$ 0.1 & 0.0 $\pm$ 0.1 & 0.0 $\pm$ 0.1  \\
    $k_v$ & 2.58 $\pm$ 1.18 & 1.39 $\pm$ 0.585 & 0.834 $\pm$ 0.245 & 0.645 $\pm$ 0.172 & 0.449 $\pm$ 0.124  \\
    $a_v$ & 0.18 $\pm$ 0.0733 & 0.26 $\pm$ 0.0732 & 0.389 $\pm$ 0.067 & 0.406 $\pm$ 0.0586 & 0.338 $\pm$ 0.043  \\
    $b_v$ & 1.65 $\pm$ 0.132 & 1.68 $\pm$ 0.125 & 1.74 $\pm$ 0.108 & 1.7 $\pm$ 0.0905 & 1.55 $\pm$ 0.0616  \\
    $k_p$ & 12.7 $\pm$ 0.922 & 13.8 $\pm$ 1.23 & 17.6 $\pm$ 2.59 & 22.4 $\pm$ 5.45 & 29.7 $\pm$ 13.3  \\
    $\chi^{2}$ & 39.5 & 39.2 & 38.8 & 39.9 & 49.1  \\
    Reduced $\chi^{2}$ & 0.527 & 0.522 & 0.517 & 0.532 & 0.655  \\
    \hline\hline
    \multicolumn{5}{c}{Fit D0} \\
    \hline\hline
    $z$ & 4.0 & 3.6 & 3.0 & 2.6 & 2.0  \\
    \hline
    $b_{\alpha}$ & -0.349 $\pm$ 0.00558 & -0.283 $\pm$ 0.00454 & -0.197 $\pm$ 0.00298 & -0.145 $\pm$ 0.00201 & -0.0794 $\pm$ 0.000957  \\
    $\beta_{\alpha}$ & 0.893 $\pm $ 0.05 & 1.08 $\pm $ 0.0515 & 1.4 $\pm $ 0.054 & 1.58 $\pm $ 0.0556 & 1.7 $\pm $ 0.0563  \\
    $k_{nl}$ & 0.456 $\pm$ 0.0166 & 0.551 $\pm$ 0.02 & 0.59 $\pm$ 0.0166 & 0.754 $\pm$ 0.017 & 0.803 $\pm$ 0.0113  \\
    $a_{nl}$ & 0.589 $\pm$ 0.00921 & 0.621 $\pm$ 0.0105 & 0.706 $\pm$ 0.00923 & 0.821 $\pm$ 0.00942 & 0.981 $\pm$ 0.00716  \\
    $k_p$ & 1.4 $\pm$ 0.0584 & 1.39 $\pm$ 0.0575 & 1.04 $\pm$ 0.033 & 1.17 $\pm$ 0.0299 & 1.2 $\pm$ 0.0195  \\
    $a_p$ & 0.811 $\pm$ 0.0143 & 0.811 $\pm$ 0.015 & 0.824 $\pm$ 0.0115 & 0.934 $\pm$ 0.0114 & 1.11 $\pm$ 0.00859  \\
    $k_{v0}$ & 0.757 $\pm$ 0.0635 & 0.931 $\pm$ 0.0728 & 1.13 $\pm$ 0.0712 & 1.18 $\pm$ 0.0626 & 1.02 $\pm$ 0.0416  \\
    $a_{v0}$ & 1.65 $\pm$ 0.115 & 1.69 $\pm$ 0.112 & 1.76 $\pm$ 0.0947 & 1.72 $\pm$ 0.0782 & 1.56 $\pm$ 0.0514  \\
    $k_{v1}$ & 0.545 $\pm$ 0.0712 & 0.659 $\pm$ 0.0885 & 0.847 $\pm$ 0.107 & 1.09 $\pm$ 0.124 & 1.3 $\pm$ 0.121  \\
    $a_{v1}$ & 0.742 $\pm$ 0.031 & 0.691 $\pm$ 0.0314 & 0.62 $\pm$ 0.0287 & 0.614 $\pm$ 0.0277 & 0.611 $\pm$ 0.0239  \\
    $\chi^{2}$ & 31.0 & 33.6 & 36.5 & 38.2 & 42.4  \\
    Reduced $\chi^{2}$ & 0.425 & 0.46 & 0.5 & 0.523 & 0.581  \\
    \hline\hline
    \end{tabular}
\end{table*}

\begin{table*}
	\centering
    \caption{Table of the fitted parameters for the splice verification comparing the true \pkt~from \textit{L320R50} and the spliced \pkt~from \textit{L320SR50}, associated to Fig.~\ref{fig:linear_param_splice_verification} with non linear parameters\com{. The $D_1$ parameterization (Equation~\ref{eq:model_p3d_nl_1}) is used here.}}
    \label{tab:results_fit_splice_verification}
    \begin{tabular}{lccccc}
    \hline\hline
    \multicolumn{5}{c}{Grid $ACCL2\_L320R50$} \\
    \hline\hline
    $z$ & 4.0 & 3.6 & 3.0 & 2.6 & 2.0  \\
    \hline
    $b_{\alpha}$ & -0.459 $\pm$ 0.00524 & -0.351 $\pm$ 0.00516 & -0.225 $\pm$ 0.0033 & -0.157 $\pm$ 0.00235 & -0.0797 $\pm$ 0.00128  \\
    $\beta_{\alpha}$ & 0.596 $\pm $ 0.0427 & 0.782 $\pm $ 0.0464 & 1.13 $\pm $ 0.0503 & 1.36 $\pm $ 0.0546 & 1.61 $\pm $ 0.0621  \\
    $q_1$ & 2.53 $\pm$ 0.0844 & 2.0 $\pm$ 0.166 & 1.22 $\pm$ 0.119 & 0.97 $\pm$ 0.1 & 0.814 $\pm$ 0.0797  \\
    $q_2$ & -1.0 $\pm$ 0.152 & -0.745 $\pm$ 0.221 & -0.228 $\pm$ 0.126 & -0.0674 $\pm$ 0.0873 & 0.0779 $\pm$ 0.0492  \\
    $k_v$ & 4.26 $\pm$ 0.922 & 2.42 $\pm$ 0.524 & 1.29 $\pm$ 0.297 & 0.906 $\pm$ 0.237 & 0.489 $\pm$ 0.174  \\
    $a_v$ & 0.238 $\pm$ 0.061 & 0.316 $\pm$ 0.0661 & 0.39 $\pm$ 0.0661 & 0.359 $\pm$ 0.0575 & 0.241 $\pm$ 0.0382  \\
    $b_v$ & 1.71 $\pm$ 0.128 & 1.71 $\pm$ 0.118 & 1.71 $\pm$ 0.0978 & 1.63 $\pm$ 0.0791 & 1.41 $\pm$ 0.0507  \\
    $k_p$ & 11.4 $\pm$ 0.385 & 12.0 $\pm$ 0.648 & 12.5 $\pm$ 0.754 & 12.6 $\pm$ 0.807 & 11.5 $\pm$ 0.662  \\
    $\chi^{2}$ & 39.4 & 39.6 & 40.3 & 39.7 & 44.9  \\
    Reduced $\chi^{2}$ & 0.448 & 0.45 & 0.459 & 0.451 & 0.51  \\
    \hline\hline
    \multicolumn{5}{c}{Grid $ACCL2\_L320SR50$} \\
    \hline\hline
    $z$ & 4.0 & 3.6 & 3.0 & 2.6 & 2.0  \\
    \hline
    $b_{\alpha}$ & -0.464 $\pm$ 0.0058 & -0.352 $\pm$ 0.00567 & -0.224 $\pm$ 0.0035 & -0.155 $\pm$ 0.0024 & -0.0777 $\pm$ 0.00121  \\
    $\beta_{\alpha}$ & 0.592 $\pm $ 0.0447 & 0.789 $\pm $ 0.0488 & 1.14 $\pm $ 0.052 & 1.36 $\pm $ 0.0554 & 1.59 $\pm $ 0.0605  \\
    $q_1$ & 2.46 $\pm$ 0.0887 & 2.03 $\pm$ 0.174 & 1.3 $\pm$ 0.121 & 1.09 $\pm$ 0.0984 & 0.963 $\pm$ 0.0739  \\
    $q_2$ & -1.0 $\pm$ 0.14 & -0.821 $\pm$ 0.222 & -0.326 $\pm$ 0.122 & -0.158 $\pm$ 0.0827 & -0.0162 $\pm$ 0.0444  \\
    $k_v$ & 3.82 $\pm$ 0.851 & 2.31 $\pm$ 0.482 & 1.41 $\pm$ 0.27 & 1.16 $\pm$ 0.22 & 0.868 $\pm$ 0.17  \\
    $a_v$ & 0.255 $\pm$ 0.0631 & 0.333 $\pm$ 0.0657 & 0.426 $\pm$ 0.0652 & 0.409 $\pm$ 0.056 & 0.315 $\pm$ 0.0371  \\
    $b_v$ & 1.73 $\pm$ 0.132 & 1.69 $\pm$ 0.117 & 1.71 $\pm$ 0.0978 & 1.61 $\pm$ 0.0788 & 1.4 $\pm$ 0.0513  \\
    $k_p$ & 11.7 $\pm$ 0.418 & 12.4 $\pm$ 0.7 & 13.0 $\pm$ 0.83 & 13.3 $\pm$ 0.919 & 12.9 $\pm$ 0.877  \\
    $\chi^{2}$ & 53.5 & 63.1 & 83.3 & 98.3 & 111.0  \\
    Reduced $\chi^{2}$ & 0.608 & 0.717 & 0.946 & 1.12 & 1.26  \\
    \hline\hline
    \end{tabular}
\end{table*}

\begin{table*}
	\centering
    \caption{Table of the fitted parameters for our the spliced \pkt~from \textit{L640SR50} and \textit{L640SR25} at all redshifts, associated to Fig.~\ref{fig:linear_param_splice} with non linear parameters\com{. The $D_1$ parameterization (Equation~\ref{eq:model_p3d_nl_1}) is used here.}}
    \label{tab:results_fit_splice}
    \begin{tabular}{lccccc}
    \hline\hline
    \multicolumn{5}{c}{Grid $ACCL2\_L640SR50$} \\
    \hline\hline
    $z$ & 4.0 & 3.6 & 3.0 & 2.6 & 2.0  \\
    \hline
    $b_{\alpha}$ & -0.464 $\pm$ 0.0054 & -0.357 $\pm$ 0.00409 & -0.229 $\pm$ 0.00252 & -0.158 $\pm$ 0.00172 & -0.0797 $\pm$ 0.000853  \\
    $\beta_{\alpha}$ & 0.639 $\pm $ 0.0362 & 0.811 $\pm $ 0.0371 & 1.14 $\pm $ 0.0388 & 1.34 $\pm $ 0.0407 & 1.56 $\pm $ 0.0435  \\
    $q_1$ & 2.52 $\pm$ 0.208 & 1.86 $\pm$ 0.167 & 1.13 $\pm$ 0.114 & 0.928 $\pm$ 0.0914 & 0.833 $\pm$ 0.0664  \\
    $q_2$ & -0.961 $\pm$ 1.77 & -0.5 $\pm$ 0.263 & -0.13 $\pm$ 0.145 & -0.0357 $\pm$ 0.097 & 0.0437 $\pm$ 0.0501  \\
    $k_v$ & 2.99 $\pm$ 1.18 & 1.67 $\pm$ 0.547 & 1.07 $\pm$ 0.264 & 0.898 $\pm$ 0.206 & 0.685 $\pm$ 0.152  \\
    $a_v$ & 0.133 $\pm$ 0.0556 & 0.205 $\pm$ 0.0604 & 0.336 $\pm$ 0.067 & 0.352 $\pm$ 0.0607 & 0.289 $\pm$ 0.0409  \\
    $b_v$ & 1.63 $\pm$ 0.134 & 1.61 $\pm$ 0.128 & 1.66 $\pm$ 0.113 & 1.62 $\pm$ 0.0952 & 1.47 $\pm$ 0.0635  \\
    $k_p$ & 10.9 $\pm$ 0.806 & 11.1 $\pm$ 0.855 & 12.0 $\pm$ 1.12 & 12.5 $\pm$ 1.31 & 12.3 $\pm$ 1.28  \\
    $\chi^{2}$ & 50.0 & 54.0 & 60.7 & 68.4 & 84.7  \\
    Reduced $\chi^{2}$ & 0.463 & 0.5 & 0.562 & 0.634 & 0.784  \\
    \hline\hline
    \multicolumn{5}{c}{Grid $ACCL2\_L640SR25$} \\
    \hline\hline
    $z$ & 4.0 & 3.6 & 3.0 & 2.6 & 2.0  \\
    \hline
    $b_{\alpha}$ & -0.408 $\pm$ 0.00499 & -0.318 $\pm$ 0.00376 & -0.208 $\pm$ 0.00232 & -0.147 $\pm$ 0.00159 & -0.0761 $\pm$ 0.000808  \\
    $\beta_{\alpha}$ & 0.902 $\pm $ 0.0398 & 1.1 $\pm $ 0.0404 & 1.42 $\pm $ 0.0417 & 1.6 $\pm $ 0.0431 & 1.75 $\pm $ 0.0452  \\
    $q_1$ & 2.5 $\pm$ 0.178 & 1.84 $\pm$ 0.139 & 1.13 $\pm$ 0.094 & 0.929 $\pm$ 0.0753 & 0.821 $\pm$ 0.0565  \\
    $q_2$ & -0.866 $\pm$ 0.282 & -0.534 $\pm$ 0.194 & -0.17 $\pm$ 0.106 & -0.0609 $\pm$ 0.0709 & 0.0388 $\pm$ 0.0381  \\
    $k_v$ & 2.2 $\pm$ 0.504 & 1.56 $\pm$ 0.297 & 1.1 $\pm$ 0.178 & 0.909 $\pm$ 0.145 & 0.612 $\pm$ 0.11  \\
    $a_v$ & 0.23 $\pm$ 0.048 & 0.331 $\pm$ 0.053 & 0.45 $\pm$ 0.0559 & 0.442 $\pm$ 0.0491 & 0.328 $\pm$ 0.0327  \\
    $b_v$ & 1.68 $\pm$ 0.108 & 1.73 $\pm$ 0.104 & 1.74 $\pm$ 0.0865 & 1.69 $\pm$ 0.0722 & 1.49 $\pm$ 0.0491  \\
    $k_p$ & 15.1 $\pm$ 1.25 & 16.3 $\pm$ 1.57 & 18.6 $\pm$ 2.41 & 19.7 $\pm$ 2.91 & 17.6 $\pm$ 2.18  \\
    $\chi^{2}$ & 58.3 & 59.5 & 63.2 & 74.0 & 107.0  \\
    Reduced $\chi^{2}$ & 0.52 & 0.531 & 0.564 & 0.661 & 0.953  \\
    \hline\hline
    \end{tabular}
\end{table*}


\bsp	
\label{lastpage}
\end{document}